\documentclass{emulateapj}
\usepackage{graphicx}
\usepackage{epstopdf}
\usepackage{amsmath}
\usepackage{amsfonts}
\usepackage{amssymb}

\shorttitle{Variability in Optical Spectra of $\epsilon$ Ori}  

\begin{document}

\title{Variability in Optical Spectra of $\epsilon$ Orionis} 

\author{Gregory B. Thompson}
\affil{Department of Physics, Adrian College, Adrian, MI 49221}
\email{gthompson@adrian.edu}
\author{Nancy D. Morrison}
\affil{Ritter Astrophysical Research Center, Department of Physics and Astronomy, University of Toledo, 2801 W. Bancroft, Toledo, OH 43606}
\email{nmorris@utnet.utoledo.edu}

%######################################################

\begin{abstract}

We present the results of a time-series analysis of 130 \'{e}chelle spectra of $\epsilon$ Ori (B0 Ia), acquired over seven observing seasons between 1998 and 2006 at Ritter Observatory.  The equivalent widths of H$\alpha$ (net) and He I $\lambda$5876 were measured and radial velocities were obtained from the central absorption of He I $\lambda$5876.  Temporal variance spectra (TVS) revealed significant wind variability in both H$\alpha$ and He I $\lambda$5876.  The He I TVS have a double-peaked profile consistent with radial velocity oscillations.  A periodicity search was carried out on the equivalent width and radial velocity data, as well as on wavelength-binned spectra.  This analysis has revealed several periods in the variability with time scales of \mbox{2-7 d}.  Many of these periods exhibit sinusoidal modulation in the associated phase diagrams.  Several of these periods were present in both H$\alpha$ and He I, indicating a possible connection between the wind and the photosphere.  Due to the harmonic nature of these periods, stellar pulsations may be the origin of some of the observed variability.  Periods on the order of the rotational period were also detected in the He I line in the 98-99 season and in both lines during the 04-05 season.  These periods may indicate rotational modulation due to structure in the wind.

\end{abstract}


\section{INTRODUCTION}

Early-type supergiant stars are the most luminous stars in the universe, and their intense radiation is responsible for driving their powerful stellar winds.  Absorption of UV photons by metal ions (e.g. N IV, O IV, Fe-group elements) transfers momentum to the ions.  Winds driven by this mechanism are known as line-driven winds.  In the 1970's it was realized that, if the momentum gained by the metal ions could be shared with the hydrogen and helium ions, which are more abundant, then this process could lead to significant mass loss (Lucy \& Solomon, 1970; Castor, Abbott \& Klein, 1975).  

The global properties of line-driven winds are successfully described by a spherically symmetric, homogeneous model wind with a velocity that increases radially outward.  This model leads to the classic P Cygni profiles indicative of mass loss.  An example of a P Cygni profile for $\epsilon$ Ori can be seen in the spectrum of 2002 February 6 shown in Figure \ref{fig1}.  However, some observations cannot be explained by this simple model.  The observed X-ray emission from hot stars is believed to be due to shocks within the wind, which may be due to the inherent line-driven instability of these winds, first described by Lucy \& Solomon (1970).  The observed electron-scattering wings of recombination lines in Wolf-Rayet stars are too weak to be matched by synthetic line profiles unless wind clumping is included in the models (Hillier (1991).  Finally, line profile variability is observed in the spectra of early-type stars.  This variability, typically observed in UV and optical lines, may be related to structure in the winds, stellar pulsations, or magnetic activity.   

Line profile variability in OB stars was firmly established with observations from the IUE satellite in unsaturated P Cygni profiles in the ultraviolet.  The emission component is fairly constant while absorption features are superimposed on the broad, blue absorption troughs.  Commonly referred to as discrete absorption components (DACs), these features typically move blueward across the profile and often recur at regular intervals on the order of the estimated rotational period.  Their cyclical nature is thought to be due to co-rotating interaction regions (CIRs) (Mullan, 1984; Cranmer \& Owocki, 1996; Kaper et al. 1997).  The CIRs may form when a density enhancement at the base of the wind propagates outward and forms a spiral density pattern in the wind as the star rotates.  The mechanism responsible for the formation of the CIRs in hot stars is not known but is likely due to disturbances at the photosphere.  Cantiello et al. (2009) propose that the cause may be tied to magnetic fields arising in sub-surface convection zones and that micro-turbulence due to these convection zones may be responsible for inducing wind clumping near the photosphere.

%Ha0102a Figure Set 1
\begin{figure*}[ht]
\figurenum{1}
\begin{center}
\includegraphics*[width= 0.93\linewidth]{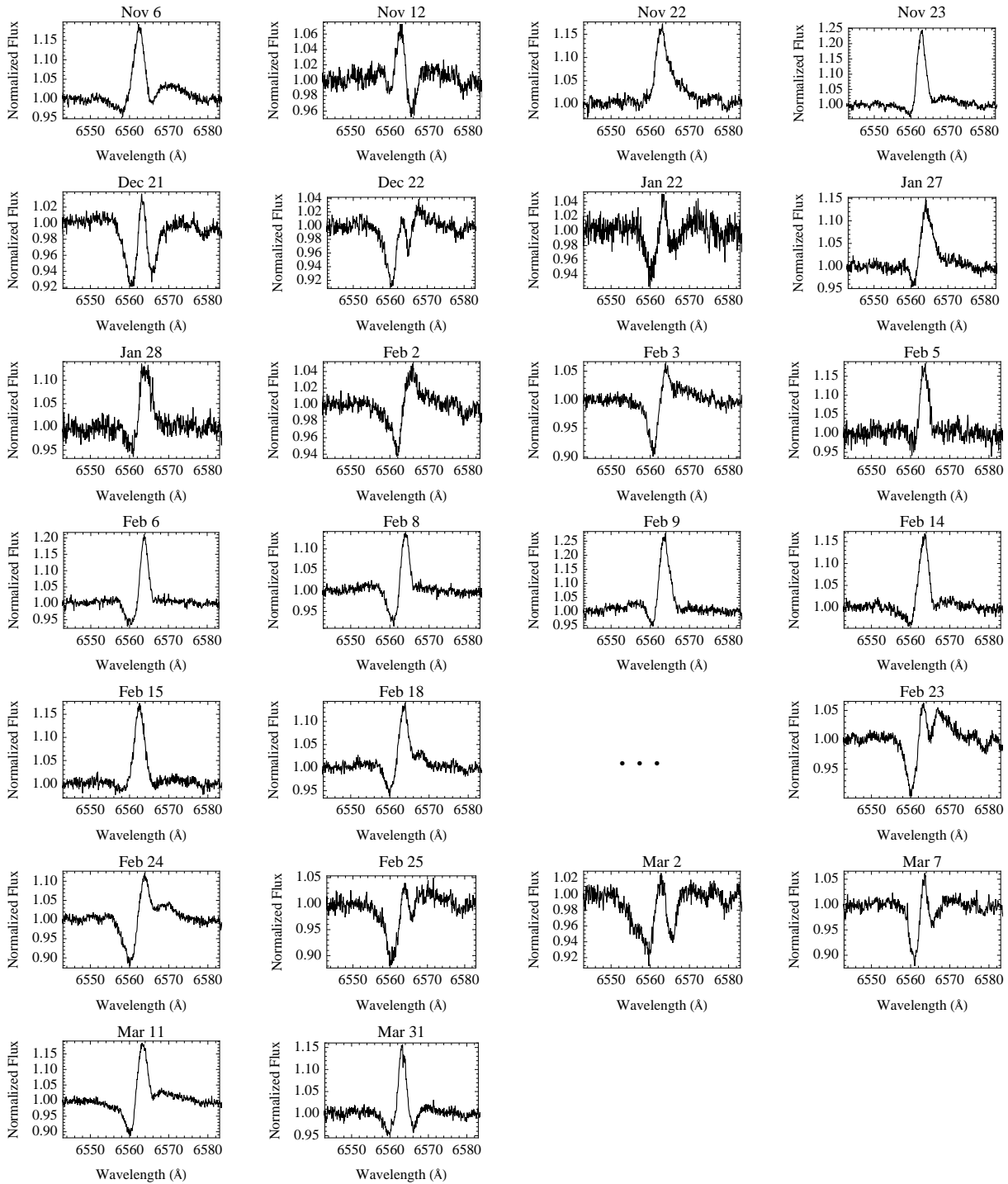}
\caption[2001-2002 H$\alpha$ Spectra]{The H$\alpha$ spectra for the 2001-2002 observing season.  UT dates are indicated for each spectrum.  (A color version and the complete figure set (14 images) are available in the online journal.)}
\label{fig1}
\end{center}
\end{figure*}

%He0102a Figure Set 1
\begin{figure*}[ht]
\figurenum{1}
\begin{center}
\includegraphics*[width= 0.93\linewidth]{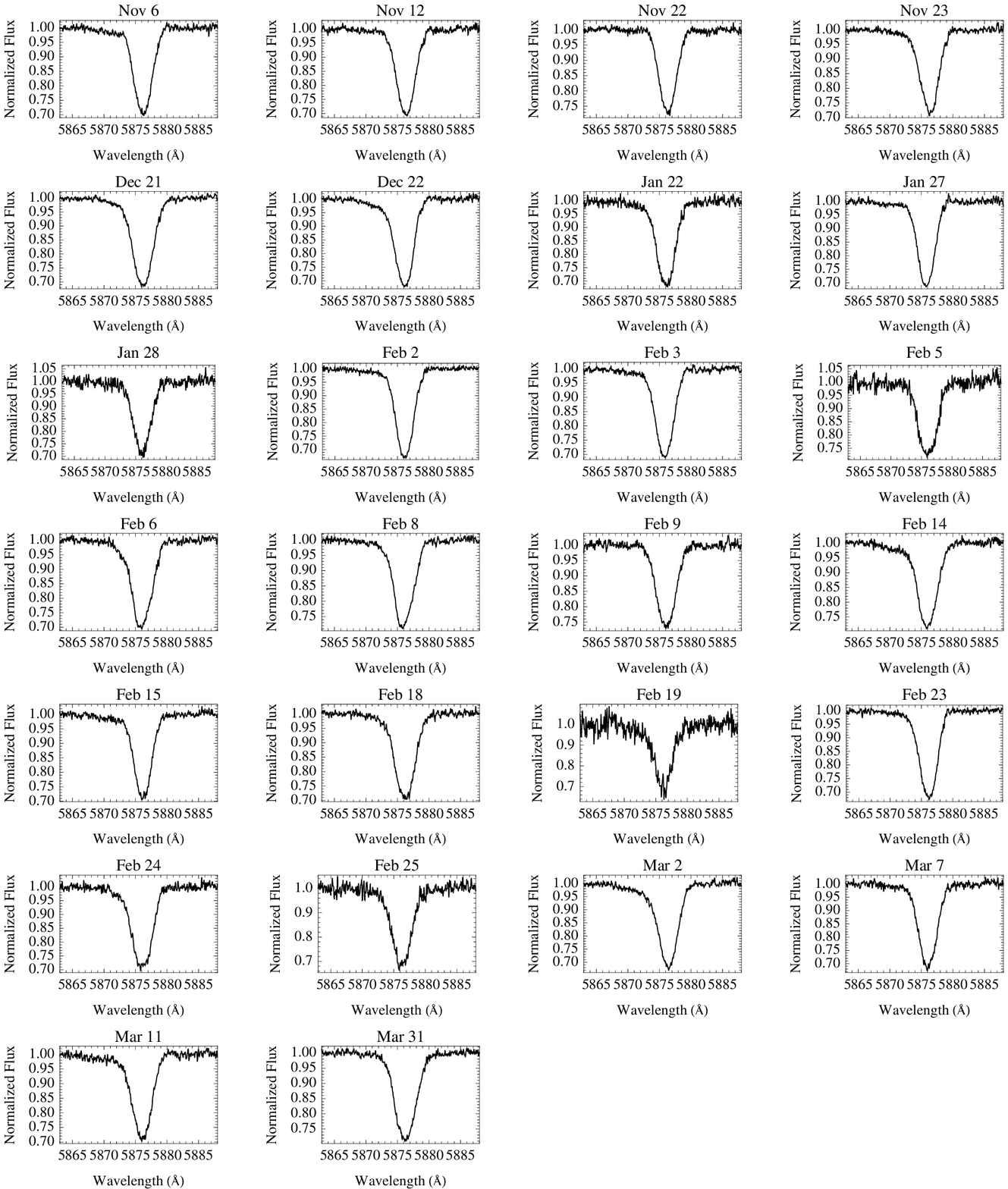}
\caption[2001-2002 He I $\lambda$5876 Spectra]{Continued.}
\label{fig1}
\end{center}
\end{figure*}
 
The H$\alpha$ line profiles of hot massive stars are also known to be highly variable.  The wind contribution of the H$\alpha$ line, being mainly formed by recombination and therefore proportional to the square of the electron density, is formed in the dense inner wind region.  In the case of weak wind emission, the observed H$\alpha$ profiles often alternate between P Cygni, inverse P Cygni, pure emission, and double absorption.  Variability in optical photospheric lines such as He I $\lambda$5876 is also observed.  Fullerton et al. (1996) found that nearly 23 out of 30 O-stars in their sample had variable photospheric lines.  The time scales of the photospheric and wind variability are often comparable, suggesting a connection between phenomena in the two regions.  One mechanism for variability in the photospheric lines is radial or non-radial pulsation.  In addition, correlations between UV and H$\alpha$ variability (Kaper et al. 1997) and between X-ray and H$\alpha$ variability (Berghoefer et al. 1996) have been reported in O-stars.  This implies the propagation of structures or disturbances throughout the wind.  

The search for a photosphere-wind connection in massive stars is of current interest.  Stellar pulsations may affect the mass loss by coupling to the wind (Townsend 2007).  In addition, Townsend (2000a, 200b) found that non-radial pulsations can ``leak'' into the base of the wind, possibly inducing the formation of wind structure that could lead to a CIR.  Kaufer et al. (2006) studied the variability of optical lines in the B0 Ib supergiant HD 64760, which is similar in spectral type to $\epsilon$ Ori (B0 Ia).  They identified three closely spaced frequencies around \mbox{5 d$^{-1}$} in photospheric lines and demonstrated that these frequencies led to a beating pattern with a period of \mbox{6.8 d}.  This \mbox{6.8 d} period was observed in photospheric lines as well as H$\alpha$.  Multi-periodic, non-radial pulsation modeling showed that three closely spaced, high-order modes could explain the beat period.  Non-radial $p$- and $g$-modes have been detected in early B-type supergiants (Lefever et al. 2007; Saio et al. 2006).  Using the $MOST$ satellite, Saio et al. detected $p$- and $g$-mode pulsations in the B2 Ib/II supergiant HD 163899 with periods ranging from $\sim$0.35 to \mbox{$\sim$47 d}.  Stellar modeling by Saio et al. of supergiants between 12 and 20 M$_\odot$ show that both $p$- and $g$-modes are excited in these stars.   

%##################################################

\section{PREVIOUS STUDIES OF $\epsilon$ ORIONIS} \label{PrevStudies}

Also known as Alnilam, $\epsilon$ Ori (HD 37128) is the middle star of Orion's Belt and is a fairly typical early B supergiant.  Some basic properties of the star are given in Table 1.  Though it is not known to be a binary, Morrell and Levato (1991) and Jarad et al. (1989) give conflicting reports of radial velocity variability for the system.  It has been reported to be moderately deficient in nitrogen, along with the other two Belt stars, $\delta$ and $\zeta$ Ori, by Walborn (1976).  $\epsilon$ Ori is one of the few early-type supergiants to have a measured radio flux density of the wind (Scuderi et al. 1998; Blomme et al. 2002).  

There have been two recent quantitative studies of early B supergiants based on UV and optical spectra and NLTE model atmosphere codes (Crowther et al. 2006; Searle et al. 2008) that included $\epsilon$ Ori.  The latter study included the effects of wind clumping.  Some NLTE derived parameters for this star reported in those two papers are listed in Table 2.  It is interesting to note the discrepancy between the $\log{g}$ and radius values found in these two studies.  Searle et al. found a larger gravity as well as a larger radius, so a major discrepancy between the masses is implied.

%TABLE 1 
\begin{deluxetable}{llllc}
\tablenum{1}
\tablecaption{\label{tab1} Basic parameters of $\epsilon$ Ori.}
\tablewidth{245pt}
\tablehead{
  \colhead{Sp. Type} &
  \colhead{$V$} &
  \colhead{Distance} &
  \colhead{$v_r$} &
  \colhead{$v \sin {i}$} \\
  \colhead{} &
  \colhead{(mag)} &
  \colhead{(pc)} &
  \colhead{(km s$^{-1}$)} &
  \colhead{(km s$^{-1}$)} }
\startdata
B0 Ia \tablenotemark{a} & 1.69 \tablenotemark{b} & $606^{+220}_{-128}$ \tablenotemark{c} & $25.90\pm0.90$ \tablenotemark{d} & 91 \tablenotemark{e}
\enddata
\tablerefs{(a) Johnson \& Morgan (1953), (b) Lee (1968), (c) van Leeuwen (2007), (d) Evans (1967), (e) Howarth et al. (1997)}
\end{deluxetable}

%TABLE 2
\begin{deluxetable*}{lcllcccc}
\centering
\tablenum{2}
\tablecaption{\label{tab2} NLTE derived parameters of $\epsilon$ Ori.}
\tablehead{
  \colhead{$T_{eff}$} &
  \colhead{$\log{g}$} &
  \colhead{$\log{L/\rm{L}_\odot}$} & 
  \colhead{$M_V$} & 
  \colhead{$R_*$/R$_\odot$} & 
  \colhead{$v_\infty$} &
  \colhead{$\xi$} &
  \colhead{Ref.} \\
  \colhead{(K)} &  
  \colhead{(cgs)} &
  \colhead{} &
  \colhead{(mag)} &
  \colhead{} &
  \colhead{(km s$^{-1}$)} &
  \colhead{(km s$^{-1}$)} &
  \colhead{} }        
\startdata
 27000 & 2.9 & 5.44 & -6.3$\pm$0.5 & 24.0 & 1910 & 12.5 & 1 \\
 27500 $\pm$ 1000 & 3.13 & 5.73 $\pm$ 0.11 & -6.89 $\pm$ 0.05 & 32.4 $\pm$ 0.75 & 1600 & 15 & 2 
\enddata
\tablerefs{(1) Crowther et al. (2006), (2) Searle et al. (2008)}
\end{deluxetable*}

Variability in $\epsilon$ Ori has been studied at multiple wavelengths including the X-ray, UV, optical, and radio regions of the spectrum.  Cassinelli et al. (1983) made simultaneous X-ray and UV observations of $\epsilon$ and $\kappa$ Ori in order to study variability in the X-ray flux and UV line profiles.  Neither star exhibited variability on time scales of 100-5000 s.  Ultraviolet observations from the IUE satellite showed variability on the time scales of $\sim$10 hours in the form of DACs and narrow absorption components (Prinja et al. 2002; Sapar \& Sapar 1998).  The IUE time series were not extensive enough to allow for a study over longer time scales.  Blomme et al. (2002) carried out an extensive radio study of $\epsilon$ Ori.  They found no evidence of radio variability and found an average 6 cm flux of $0.74 \pm 0.13$ mJy.  Blomme et al. reported flux measurements at millimeter wavelengths that are larger than what is expected for a smooth wind model.

The most extensive studies of $\epsilon$ Ori concern the variability of optical line profiles, particularly H$\alpha$.  The H$\alpha$ emission in $\epsilon$ Ori was first noted by Cherrington (1937) in which he described the line as having a ``peculiar structure.''  The variability of H$\alpha$ in $\epsilon$ Ori was studied in some detail by Ebbets (1982) as part of a survey of variability in early-type stars.  Ebbets noted that the broad emission in H$\alpha$ typically extended from -800 km s$^{-1}$ to 800 km s$^{-1}$ and the average equivalent width of H$\alpha$ for the 6 observations of $\epsilon$ Ori was -0.56 \r{A} with a range of -0.01 to -0.96 \r{A}.  The time scale for the observed variability in H$\alpha$ and the photospheric lines was 1-10 d. 

Morel et al. (2004) conducted a time-series analysis of H$\alpha$ line profile variability for 22 OB supergiants.  Their study included 21 observations of $\epsilon$ Ori obtained on 17 nights between Nov. 18, 2001 and May 7, 2002.  They conducted a periodicity search on the measured net equivalent widths of H$\alpha$ and the pixel-by-pixel H$\alpha$ time series.  Significant variability was observed in H$\alpha$ and He I $\lambda$6678, as quantified using temporal variance spectra.  On several nights multiple observations were obtained for $\epsilon$ Ori and significant changes in the H$\alpha$ profile were observed on time scales of a few hours.  Their period analysis found periods of \mbox{0.781 d} and \mbox{18.2 d} in the H$\alpha$ data.  The \mbox{18.2 d} period is on the order of the rotational period of $\epsilon$ Ori, for which Morel et al. quote a maximum of \mbox{17.6 d}.  Prinja et al. (2004) quote a maximum rotational period of \mbox{22 d}.

A more extensive time-series analysis of H$\alpha$ and photospheric lines in $\epsilon$ Ori was conducted by Prinja et al. (2004).  Their study was primarily based on 61 spectra obtained over 17 nights between Nov. 26 and Dec. 13, 1998, supplemented by archival spectra taken over several months in 1996.  In addition to H$\alpha$, their data included the absorption lines H$\beta$, H$\gamma$, He I $\lambda$4713, He I $\lambda$5876, He I $\lambda$6678, C II $\lambda$4267, and Si III $\lambda$4553.  Radial velocity variations were observed in all of the absorption lines.  The period analysis involved constructing 2D (pixel-by-pixel) periodograms using the CLEAN algorithm.  They report periods of 1.9, 6.6 and \mbox{9.7 d}, each identified in H$\alpha$, H$\beta$, and He I $\lambda$6678.  Finally, they carried out a non-radial pulsation study of the \mbox{1.9 d} period to see if non-radial pulsations could replicate the observations.  The authors concluded that non-radial pulsations are not sufficient to explain the observed variability for the \mbox{1.9 d} period. 

This paper examines the line profile variability of the H$\alpha$ and He I $\lambda$5876 lines of $\epsilon$ Ori.  The remainder of this paper is organized in the following manner.  Section \ref{sec3} describes the instrumentation, observations, and data reduction.  Section \ref{sec4} discusses the temporal variance spectra and the equivalent width and radial velocity measurements.  Details of the line profile variability are described in Section \ref{sec5}.  A discussion of the techniques used in the period search is presented in Section \ref{sec6}.  The results of the period search are presented in Section \ref{sec7}.  Finally, a summary of the findings and implications for future work are given in Section \ref{sec8}. 

%####################################################

%TABLE 3
\def\arraystretch{1.2}
\begin{deluxetable}{lccc}
\tablenum{3}
\tablecaption{\label{tab3} Summary of Observations.}
\tablewidth{245pt}
\tablehead{
  \colhead{Season} &
  \colhead{Number of Spectra} &
  \colhead{$\Delta$$t$} &
  \colhead{$T$} \\
  \colhead{} &
  \colhead{} &
  \colhead{(d)} &
  \colhead{(d)} }    
\startdata
1998-1999 & 15 & 8.2 & 115 \\ 
2000-2001 & 14 & 12.5 & 163 \\ 
2001-2002 & 26 & 6.0 & 145 \\ 
2002-2003 & 21 & 6.4 & 128 \\ 
2003-2004 & 10 & 16.5 & 149 \\ 
2004-2005 & 24 & 8.9 & 205 \\ 
2005-2006 & 22 & 9.8 & 196 
\enddata
\tablecomments{$\Delta$$t$ refers to the average interval between observations and $T$ refers to the total span of the time series.}
\end{deluxetable}

\section{INSTRUMENTATION, OBSERVATIONS, AND DATA REDUCTION} \label{sec3}
      
One hundred and thirty spectra of $\epsilon$ Ori were obtained at Ritter Observatory using the 1-meter Ritchey-Chr\'{e}tien telescope and fiber-fed \'{e}chelle spectrograph.  Target acquisition and manual guiding were done with a SBIG ST-9 CCD camera.  The detector for the spectrograph was a front-illuminated EEV CCD in a Wright Instruments camera system.  The CCD had 1200 $\times$ 800 pixels with physical dimensions of 22.5 $\times$ 22.5 $\mu$m.  The detector was cooled with liquid nitrogen to an operating temperature of 140 K.  The entrance slit of the spectrograph projected to a width of 4.3 pixels and corresponded to a resolving power R = 26,000.  In its usual configuration, nine partial \'{e}chelle orders were included on the chip with H$\alpha$ falling in the center of the lowest-numbered order on the CCD.  The included \'{e}chelle orders, each spanning $\sim$70 \r{A}, covered the spectral range from 5285 to 6594 \r{A}.
      
In this paper, spectra of $\epsilon$ Ori from seven observing ``seasons'' between 1998 and 2006 are analyzed.  There were no observations during the 1999-2000 season.  In Toledo, Ohio, an observing season for Orion spans roughly from September to April.  A typical season from this data set includes $\sim$20 spectra, each with a signal-to-noise ratio (S/N) of $\sim$100 or better per pixel in the continuum.  The average interval between observations is $\sim$10 nights.  Prominent features in the Ritter spectra of $\epsilon$ Ori include the H$\alpha$ line, the He I 5876 \r{A} D3 multiplet, as well as interstellar Na I D lines.  Weak C II lines at 6578 and 6583 \r{A} are visible in a few spectra.  This work examines the line profile variability of H$\alpha$ and He I $\lambda$5876.  The time series are summarized in Table 3.

The raw  \'{e}chelle spectra were reduced in IRAF\footnote{IRAF is distributed by the National Optical Astronomy Observatories, which are operated by the Association of Universities for Research in Astronomy, Inc., under cooperative agreement with the National Science Foundation.} with the observatory's pipeline script.  The standard reductions included (in this order) bias subtraction, \'{e}chelle order extraction, flat fielding, and wavelength calibration.  The wavelengths were calibrated using comparison spectra from a Th-Ar discharge lamp.  The rms of the calibration fit is typically between 0.001 - 0.005 \r{A} (Mulliss, 1996).  Further operations on spectra produced by the pipeline included the removal of telluric lines, corrections for the motion of the Earth about the Sun, and normalization to the continuum.

Beginning with data taken in the spring of 2002, there was a problem with the wavelength calibration coming out of the standard pipeline.  It is suspected that the problem began after the observatory replaced the Th-Ar lamp.  As a result, beginning with the 2002-2003 season, the spectra had to be re-calibrated.  The re-calibration used known telluric water lines that appear in the spectra in same way that Th-Ar lines were used in the original calibration.  The wavelengths of the water lines used were measured in the telluric spectrum provided in FITS format by Hinkle et al. (2000).  The water lines were identified in the spectra using the IRAF task \emph{identify} and the known wavelengths were fitted to the observed spectrum.  The dispersion solution was then applied to the spectrum to correct the wavelength scale.  This procedure was carried out separately for the two apertures containing H$\alpha$ and He I $\lambda$5876.  The He I $\lambda$5876 and H$\alpha$ spectra from the 2001-2002 season are shown in Figures \ref{fig1} and \ref{fig1}, respectively.  Spectra from all seven observing seasons are available in the online version of the journal. 

%###############################################

%Table 4 stub
\def\arraystretch{1.1}
\begin{deluxetable}{cccc}
\tablenum{4}
\tablecaption{\label{tab4} Equivalent Width and He I Radial Velocity Data.}
\tablewidth{245pt}
\tablehead{
  \colhead{HJD-2450000} &
  \colhead{$W_{H\alpha}$} &
  \colhead{$W_{HeI}$} &
  \colhead{$V_r$} \\
  \colhead{} &
  \colhead{(\r{A})} &
  \colhead{(\r{A})} &
  \colhead{(km s$^{-1}$)} }    
\startdata
1149.803	&	-0.3801	&	1.139	&	11.8	\\
1163.736	&	-0.1594	&	1.080	&	8.53	\\
1208.596	&	-0.0179	&	1.125	&	11.7	\\
1214.670	&	-0.0530	&	1.215	&	7.20	\\
1220.648	&	-0.5581	&	1.010	&	0.52	
\enddata
\tablecomments{This table is available in its entirety in machine-readable and Virtual Observatory (VO) forms in the online journal. A portion is shown here for guidance regarding its form and content.}
\end{deluxetable}
      
\section{THE TVS, EQUIVALENT WIDTHS, AND RADIAL VELOCITY MEASUREMENTS} \label{sec4}
\subsection{The Temporal Variance Spectra}
In order to quantify the line profile variability, temporal variance spectra were calculated for H$\alpha$ and He I line for each season.  The temporal variance spectrum was first outlined in Fullerton (1990) and Fullerton, Gies and Bolton (1996).  The advantage of the technique is that it allows one to determine the level of statistical significance of detected line profile variability.  The TVS method accounts for spectrum-to-spectrum and pixel-to-pixel differences in data quality.  
           
\begin{figure*}[htb]
\figurenum{2}
\begin{center}
\includegraphics*[width=0.75\linewidth]{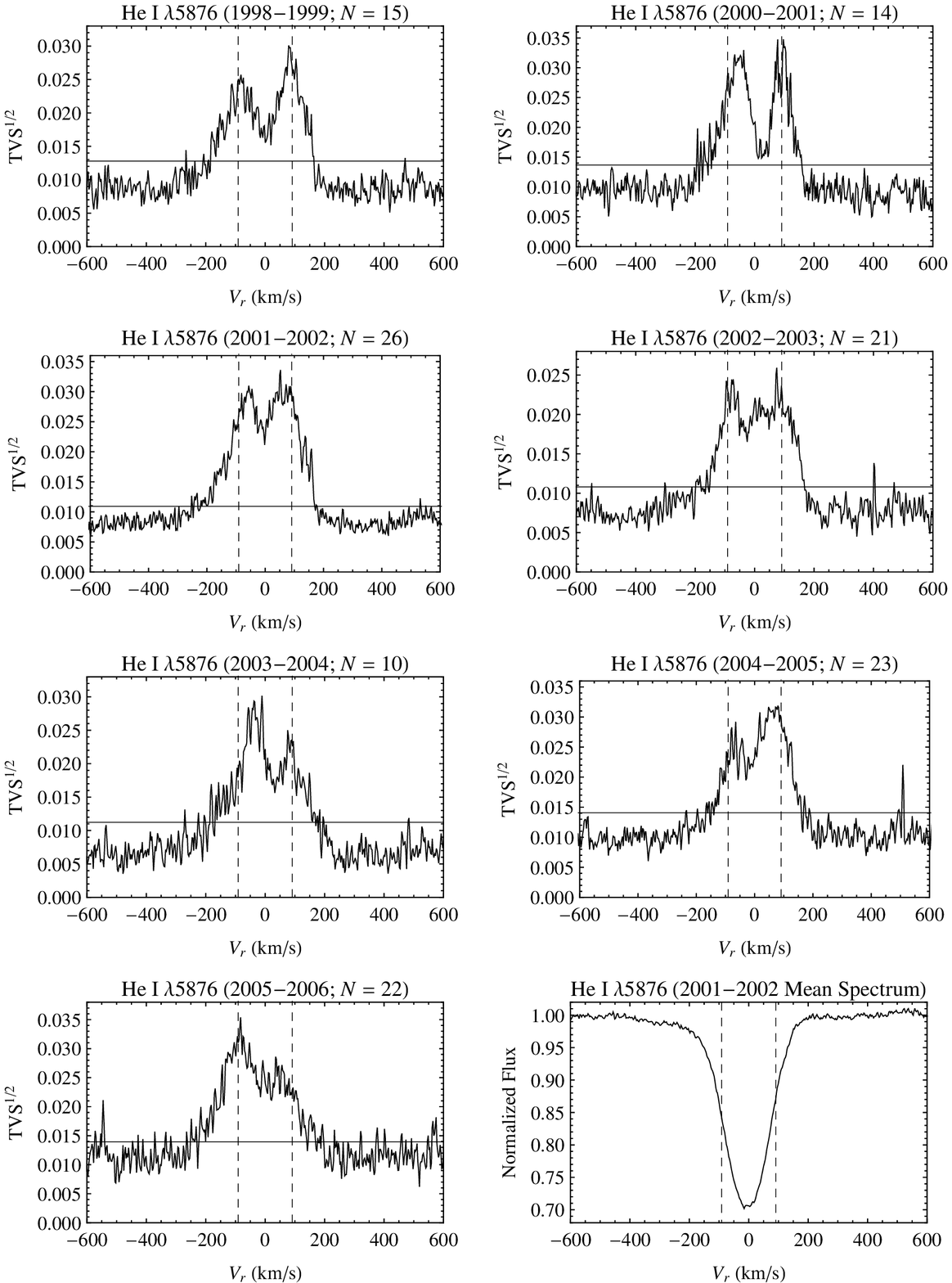}
\caption[H$\alpha$ TVS]{The temporal variance spectra, plotted as TVS$^{1/2}$, for the He I spectra from all seasons.  The horizontal line indicates the level of significant variability at the 99\% confidence level and the vertical dashed lines mark $v\sin{i}$ at $\pm$91 km s$^{-1}$.  The velocity scale has been converted to the stellar rest frame.  For reference, the average spectrum from 2001-2002 is shown in the bottom right panel.}
\label{fig2}
\end{center}
\end{figure*}

\begin{figure*}[htb]
\figurenum{3}
\begin{center}
\includegraphics*[width=0.75\linewidth]{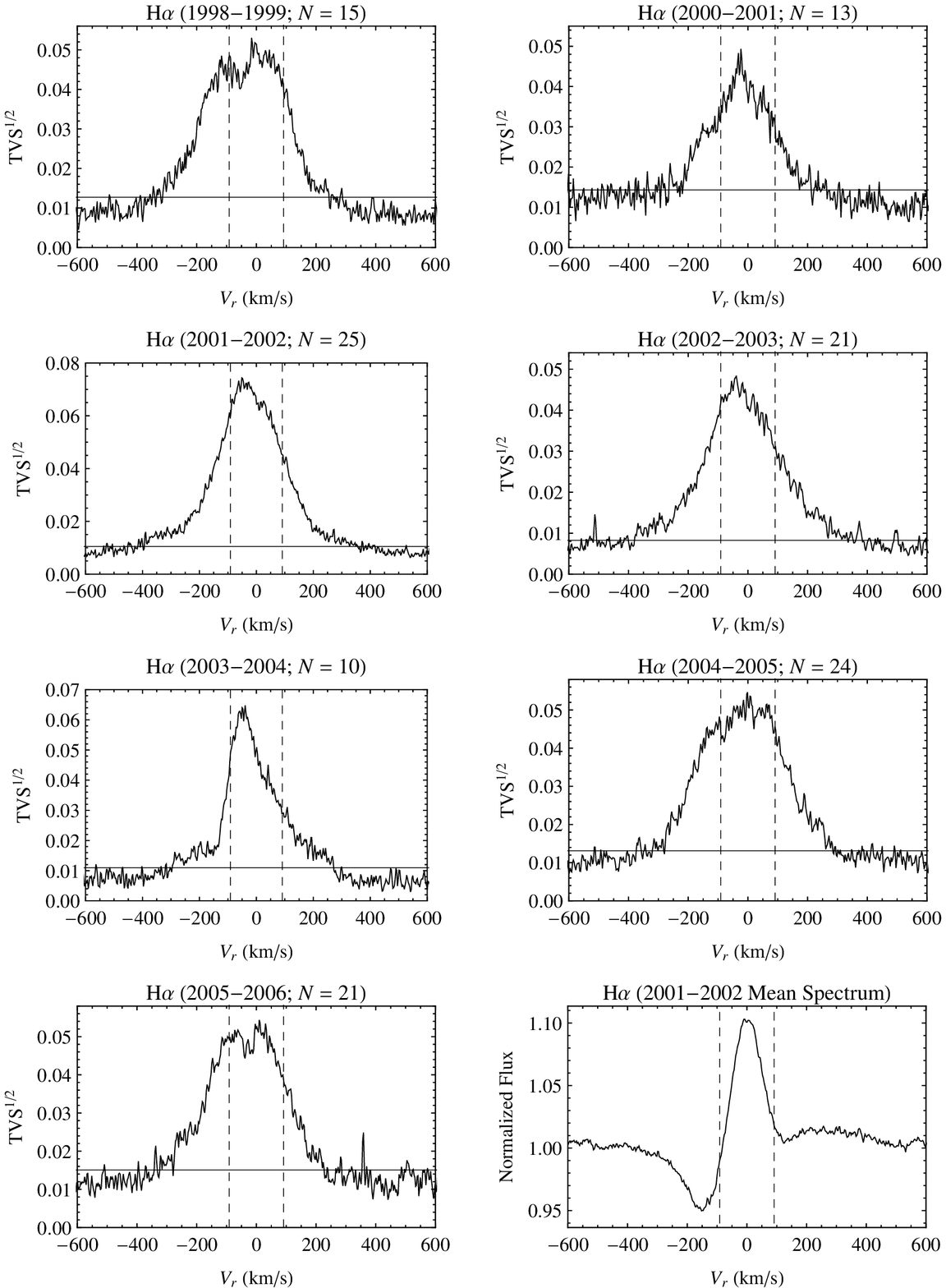}
\caption[H$\alpha$ TVS]{The temporal variance spectra, plotted as TVS$^{1/2}$, for the H$\alpha$ spectra from all seasons.  The horizontal line indicates the level of significant variability at the 99\% confidence level and the vertical dashed lines mark $v\sin{i}$ at $\pm$91 km s$^{-1}$.  The velocity scale has been converted to the stellar rest frame.  For reference, the average spectrum from 2001-2002 is shown in the bottom right panel.}
\label{fig3}
\end{center}
\end{figure*}

The TVS of a time series of $N$ spectra has the same statistical distribution as ${\sigma}^2\chi_{N-1}^2$, where $\chi_{N-1}^2$ is the chi-squared distribution with $N-1$ degrees of freedom.  Thus, it is possible to calculate the significance level for a given value of the TVS from the appropriate ${\sigma}^2\chi_{N-1}^2$ distribution.  It is then easy to determine the wavelength or velocity range over which significant variability occurs.  The He I and H$\alpha$ temporal variance spectra for all seasons are shown in Figures \ref{fig2} and \ref{fig3}.  The TVS$^{1/2}$ is plotted because it scales linearly with the spectral standard deviations.  

%HeTVS fig2

%HaTVS fig3

\subsection {Equivalent Width Measurements}
The region of significant variability at the 99\% confidence level was identified from the TVS for each season for both H$\alpha$ and the He I 5876 \r{A} triplet.  The range of significant variability was then used in defining the wavelength region over which the equivalent width, $W_\lambda$, was measured.  The goal was to define a consistent wavelength range for each line that would be used for all $W_\lambda$ measurements without excluding any significant variability.  The bluest extent of significant variability, from any of the seasons, defined the blue edge, $\lambda_b$, of the region used to measure the equivalent widths.  Similarly, the reddest extent of significant variability, from any season, defined the red edge, $\lambda_r$, of the region used to measure the equivalent widths.  This was done for both the H$\alpha$ and the He I $\lambda$5876 lines.  The equivalent width is then defined as
\begin{equation}
W_{\lambda} = \int_{\lambda_b}^{\lambda_r}(1 - F_\lambda) d\lambda.
\end{equation}
The equivalent widths of each line were measured over the ranges of 6554.25 to 6571.75 \r{A} and 5871.7 to 5879.8 \r{A} for H$\alpha$ and He I, respectively.  For H$\alpha$, the net equivalent width was measured which included both emission and absorption components.  For this reason, the $W_\lambda$ values for H$\alpha$ are small, often close to zero.  The equivalent width data for both H$\alpha$ and He I may be found in Table 4.  The errors in the equivalent width measurements were computed using the method given in Chalabaev \& Maillard (1983).  For He I, the errors were $\sim$0.035 \r{A} or $\sim$3\%.  For H$\alpha$, the errors were $\sim$0.089 \r{A}.

\subsection{Radial Velocity Measurements}
Since He I $\lambda$5876 is primarily a photospheric absorption line the radial velocity of the absorption core was measured to search for variability.  Absorption from the wind is present in the wings of the line, particularly at the blue edge.  However, the line is symmetrical at normalized flux levels below $\sim$0.9, suggesting a negligible wind contribution there.  Only this central region of the line, from the flux minimum up to 0.9 in normalized flux, was used in the radial velocity determinations.  The centroid of the line core was calculated using IRAF and the wavelength was converted to a radial velocity using a rest wavelength of 5875.704 \r{A}\footnote{Personal communication to N.D.M. from P. S. Conti (1978).  Obtained by a method similar to that of Conti et al. (1977).}.  This method attempts to minimize any wind contribution to the radial velocity variability.  However, in cases of particularly strong wind absorption or emission, it is possible that the wind may contribute to variability in the line core.   The radial velocity data may be found in Table 4.

The errors for the radial velocity measurements were calculated using the method of Brown (1990).  A typical value for $\delta v$ is $\sim$0.25 km s$^{-1}$.  Previous radial velocity studies using the same instrument at Ritter Observatory have found an instrumental error of 0.25 km s$^{-1}$ due to motions of the CCD relative to the spectrum.  The instrumental errors and the measurement errors were added in quadrature to produce the total error of the radial velocity data.

%#####################################

\section{CHARACTERIZING THE VARIABILITY} \label{sec5}

The spectra shown in Figures \ref{fig1} and \ref{fig1} demonstrate the variability in the H$\alpha$ and He I $\lambda$5876 lines of $\epsilon$ Ori.  The H$\alpha$ profile is highly variable, often appearing as P Cygni, nearly pure emission, or double absorption.  In contrast to H$\alpha$, the He I line is predominantly an absorption line with weak wind variability present in the wings.  Enhanced absorption is often observed in the blue wing of the line and a slight emission bump is sometimes seen at the red edge of the line.  The strength of the absorption line is fairly constant but the profile often oscillates back and forth.  Prinja et al. (2004) noted this ``swaying'' in the profiles of several absorption lines, including He I $\lambda$5876. 

\begin{figure*}[htb]
\figurenum{4}
\begin{center}
\includegraphics*[width=0.94\linewidth]{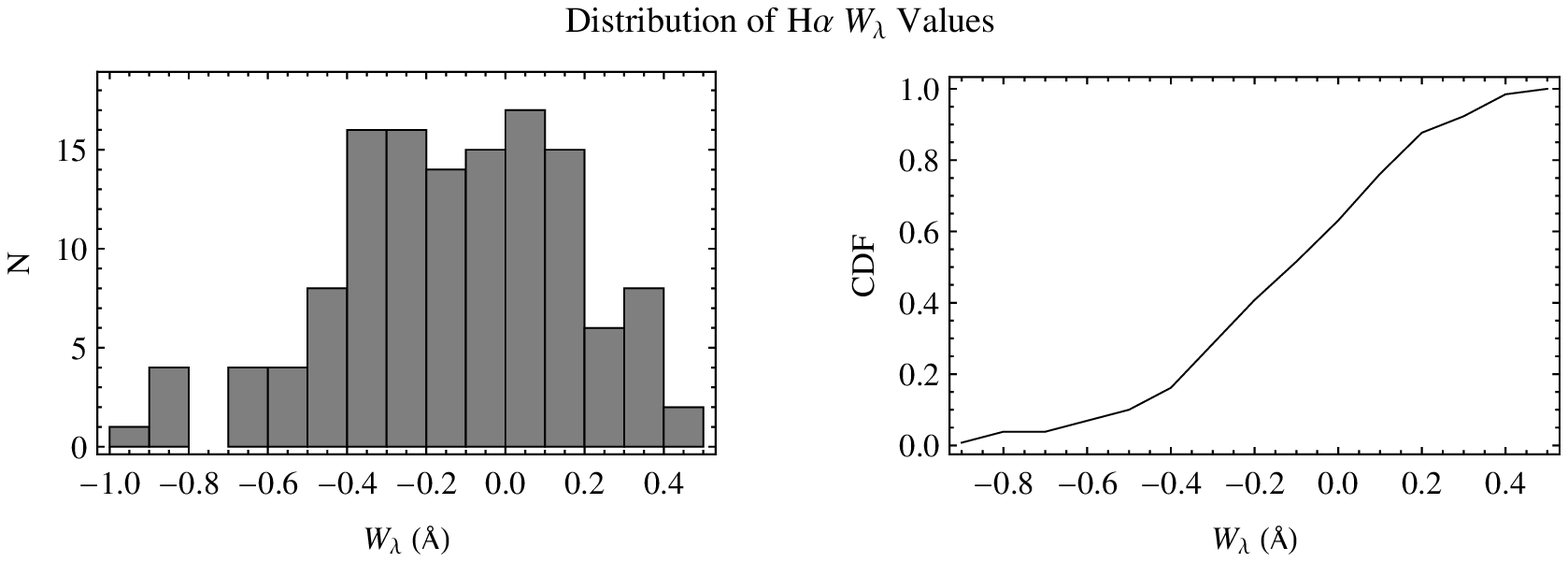}
\caption[Histogram and CDF of H$\alpha$ $W_\lambda$ Values]{The left panel shows the histogram of all H$\alpha$ equivalent width measurements.  The right panel shows the cumulative distribution function of the $W_\lambda$ values.  Both figures indicate net emission is present in most of the observations.}
\label{fig4}
\end{center}
\end{figure*}

Since it is the variable wind emission that dominates the measured H$\alpha$ equivalent widths, it is worth while to examine the distribution of the equivalent width values to gain insight into the typical wind emission.  The left frame of Fig. \ref{fig4} shows a histogram of all 130 H$\alpha$ $W_\lambda$ measurements, where the values have been put into 0.1 \r{A} bins.  The distribution is approximately Gaussian, with the exception of a slight tail extending beyond -0.6 \r{A}.  The peak of the distribution is near -0.1 \r{A}, indicating a slight preference toward net emission in the profile.  The right side of Fig. \ref{fig4} shows the cumulative distribution function of the H$\alpha$ $W_\lambda$ values.  The cumulative distribution function (CDF) indicates that net emission is observed approximately 65\% of the time.  

%HistCDF fig4

The temporal variance spectra, shown in Figures \ref{fig2} and \ref{fig3}, provide more insight into the observed variability.  The horizontal line indicates the level of significant variability at the 99\% confidence level of the ${\sigma}^2\chi_{N-1}^2$ distribution.  The vertical dashed lines mark the $v\sin{i}$ of $\pm$91 km s$^{-1}$ (Howarth et al. 1997).  The velocity scale has been converted to the stellar rest frame.  Significant variability extends beyond the $v\sin{i}$ threshold in both lines.  The range of significant variability for H$\alpha$ extends approximately -350 to 300 km s$^{-1}$ while for He I, the range is roughly -200 to 175 km s$^{-1}$.  The variability in both lines is confined to velocities much less than the terminal wind speed of 1600 km s$^{-1}$ (Searle et al. 2008).  The H$\alpha$ TVS are fairly symmetric, with the exception of the 2003-2004 season, and the peaks are shifted toward the blue.  This indicates that the bulk of the observed variability occurs in the column of wind material approaching the observer.  

The He I TVS are characterized by double peaks with the central minimum occurring near rest velocity as one would expect for a line undergoing radial velocity variations.  The double peaks are nicely contained within the $\pm$91 km s$^{-1}$ threshold, consistent with a photospheric origin for the radial velocity variations.  It should be noted that the projected rotational velocity is an upper limit, since the line broadening is likely due to macroturbulence in addition to rotation (Ryans et al. 2002).  The He I TVS tend to be slightly asymmetric with broader wings on the blue side, likely due to the wind absorption often seen on the blue edge.  

%ModelComp2 fig5
\begin{figure*}[tb]
\figurenum{5}
\begin{center}
\includegraphics*[width=0.65\linewidth]{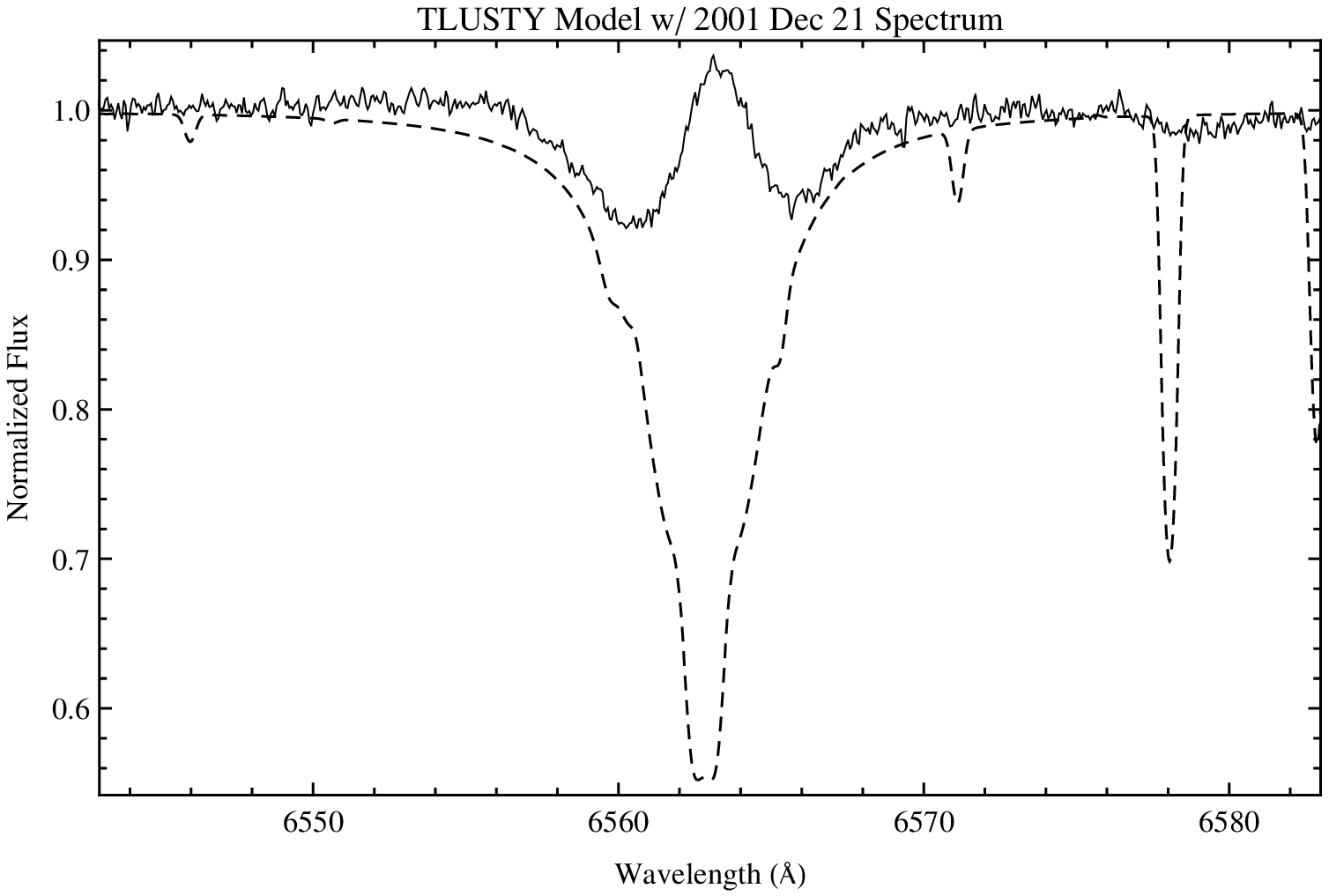}
\caption[TLUSTY Model H$\alpha$ Profile]{A TLUSTY synthetic H$\alpha$ profile (dashed) with a superimposed $\epsilon$ Ori spectrum from 2001 Dec 21 (solid).  The parameters of the model are $T_{eff} = 27000$ K, $\log{g} = 3.0$, $\xi = 10$ km s$^{-1}$, and solar metallicity.}
\label{fig5}
\end{center}
\end{figure*}  

\subsection{Physical Interpretation of the Line Profile Variability}
The H$\alpha$ photospheric absorption is heavily filled in by wind emission and photospheric variability is masked by the wind.  To better understand the underlying photospheric absorption, a TLUSTY photospheric model with a superimposed $\epsilon$ Ori H$\alpha$ spectrum is shown in Fig. \ref{fig5}.  The spectrum, from 2001 Dec 21, is an example of a minimum emission profile.  The synthetic profile is from the BSTAR2006 grid of NLTE line-blanketed model atmospheres (Lanz \& Hubeny 2007).  The model has parameters very similar to those of $\epsilon$ Ori with $T_{eff} = 27000$ K, $\log{g} = 3.0$, $\xi = 10$ km s$^{-1}$, and solar metallicity (see Table 2 and references therein).  The model profile has not been rotationally broadened.  Even when the emission is weak, it is clear that the absorption is nearly completely filled in by wind emission.  Variability in the strength of wind emission gives rise to the many observed profile types.  For example, strong emission gives rise to the emission or P Cygni profiles seen in the Nov 22 and Feb 6 spectra of Fig. \ref{fig1}.  Weaker wind emission leads to the double absorption-type profile in the Dec 21 spectrum shown in Fig. \ref{fig1}.  Since wind emission is the dominant factor in shaping the observed profile of H$\alpha$, most of the variability in the measured net equivalent widths is likely attributable to the wind.  However, in the case of constant wind emission, variability in the measured equivalent widths could be due to a variable photospheric component.

In contrast to H$\alpha$, the morphology of the He I $\lambda$5876 line is dominated by photospheric absorption.  However, weak wind features are observed as absorption in the red and blue wings and emission in the red wing.   It is important to note that the range used in measuring the He I equivalent widths (5871.7 - 5879.8 \r{A}) included a portion of these wind signatures.  The absorption in the blue wing seen in most of the spectra is fairly broad and does lie within the region of significant variability.  Wind absorption is also observed on the red edge as well, as seen in the Dec 22 spectrum of Fig. \ref{fig1}.  This absorption in the red occurs over a narrower region but also lies within the region of significant variability.  Weak emission in the red wing, as seen in the Nov 6 spectrum of Fig. \ref{fig1}, is observed much less frequently and occurs right on the edge of the region of significant variability, near 5880 \r{A}.  This emission also occurs at the edge of the region over which the equivalent widths are measured; therefore, any contribution would be very minor and may not be significant.  

\begin{figure}[htb]
\figurenum{6}
\begin{center}
\includegraphics[width=1.0\linewidth]{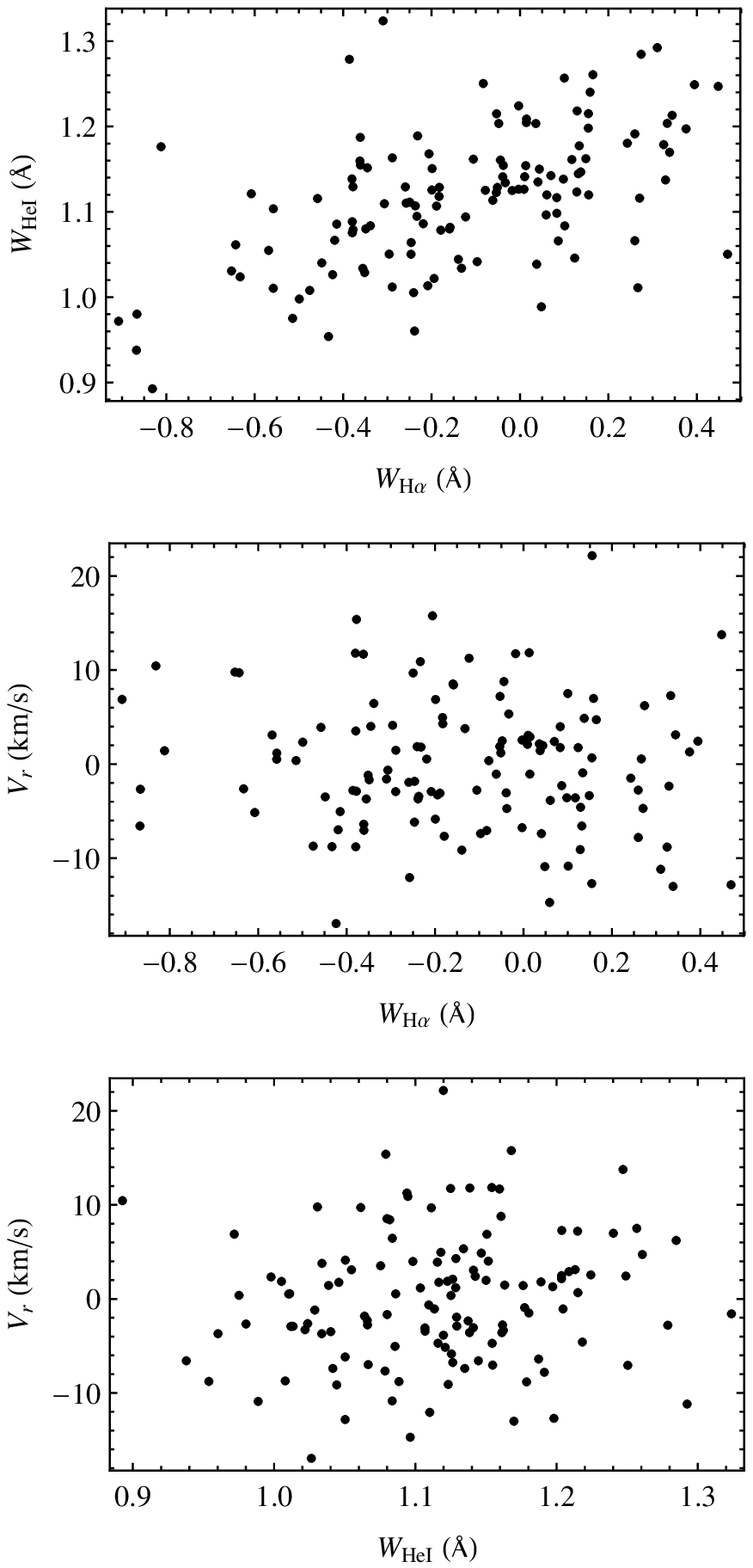}
\caption[Correlation Plots Between $W_{H\alpha}$, $W_{HeI}$, and $V_r$]{Plots of $W_{H\alpha}$ vs. $W_{HeI}$, $W_{H\alpha}$ vs. $V_r$, and $W_{HeI}$ vs. $V_r$.  The radial velocities are from the He I line.}
\label{fig6}
\end{center}
\end{figure}

Boyajian et al. (2007) observed a correlation between H$\alpha$ emission and radial velocity changes in the He I $\lambda$6678 line in the B3 Ia supergiant HD 61827.  They attribute the radial velocity variations to weak wind emission in the red wing of the He I line causing an apparent shift of line center toward the blue.  Therefore, although the core of the He I $\lambda$5876 line in $\epsilon$ Ori appears to be photospheric, the variability in the measured radial velocities could have a wind origin.  The equivalent widths of H$\alpha$ and He I, and the radial velocities were plotted against one another to look for any correlations between them.  These plots are shown in Fig. \ref{fig6}.  There does appear to be a weak, positive correlation between the equivalent widths of H$\alpha$ and He I, with a Pearson's correlation coefficient $r^2$ = 0.29.  If a connection exists between activity in the photosphere and the wind, one would expect to find such a correlation.  A variable wind component would have a smaller effect on He I than on H$\alpha$ and the small slope of 0.15 seems to support that.  There is at best a very weak correlation between either $W_{H\alpha}$ and radial velocity or $W_{HeI}$ and radial velocity, with correlation coefficients of 0.0062 and 0.014, respectively.  This suggests that the He I radial velocity measurements are not contaminated by wind emission and do originate in the photosphere.  

%CorrPlots fig6
    
%#############################################

\section{THE PERIOD SEARCH} \label{sec6}
In the present work, the most important tool for finding periodicities is the periodogram.  Two standard techniques  for handling unevenly sampled data such as these are the Lomb-Scargle periodogram (Lomb, 1976; Scargle, 1982) and the CLEAN algorithm (Roberts et al. 1987).  Another approach, outlined by Bretthorst (2001a, 2001b), combines Bayesian inference with the Lomb-Scargle (LS) periodogram.  All three methods were employed in the analysis presented here.  The period search was done on the data from each season except for 2003-2004.  That season only had 10 observations with an average time of \mbox{16.5 d} between observations and it was felt that this was insufficient sampling for a proper analysis.  A global analysis was also done on the combined data from all seasons, including 2003-2004, but no long-term trends were found.

\subsection{The Bayesian Periodogram}
Bretthorst (2001a, 2000b) showed that the Lomb-Scargle (LS) periodogram could be derived in a general way by applying Bayesian probability theory to the Lomb model (Lomb, 1976; Scargle, 1982).  For astronomical time series, assuming a sinusoidal signal, Bretthorst's algorithm yields the LS periodogram as well as the Bayesian probability of the peaks in the LS periodogram.  The algorithm assumes a Jeffreys prior which represents equal probability per decade in $f$ (Gregory, 2005).  A plot of the resulting Bayesian probability density (PDF) as a function of frequency resembles a typical periodogram but many of the smaller peaks are suppressed.  For simplicity, the plot of the Bayesian probability of the LS periodogram will be referred to as the \emph{Bayesian periodogram}.

The Lomb-Scargle and Bayesian periodograms were calculated for the equivalent width data of both H$\alpha$ and He I $\lambda$5876 as well as the radial velocities from the He I line.  The Lomb-Scargle and Bayesian periodograms were calculated using \emph{Mathematica} code adapted from files found in Gregory (2005).  Prior to calculating the periodograms, the average of each time series was subtracted.  Since the smallest interval between successive observations was \mbox{$\sim$1 d}, the frequency range was limited to $0 < f < 0.5$ d$^{-1}$, because 0.5 d$^{-1}$ is the Nyquist frequency.  The Bayesian analysis covered the frequency range $0.0001 < f < 0.5$ d$^{-1}$, with steps of $\delta f = 0.0005$.  The probability density was normalized so that
\begin{equation}
\sum^{0.5}_{f=0} PDF\times \delta f = 1
\end{equation}
where the frequency units are d$^{-1}$.

\begin{figure*}[htb]
\figurenum{7}
\begin{center}
\includegraphics*[width=0.65\linewidth]{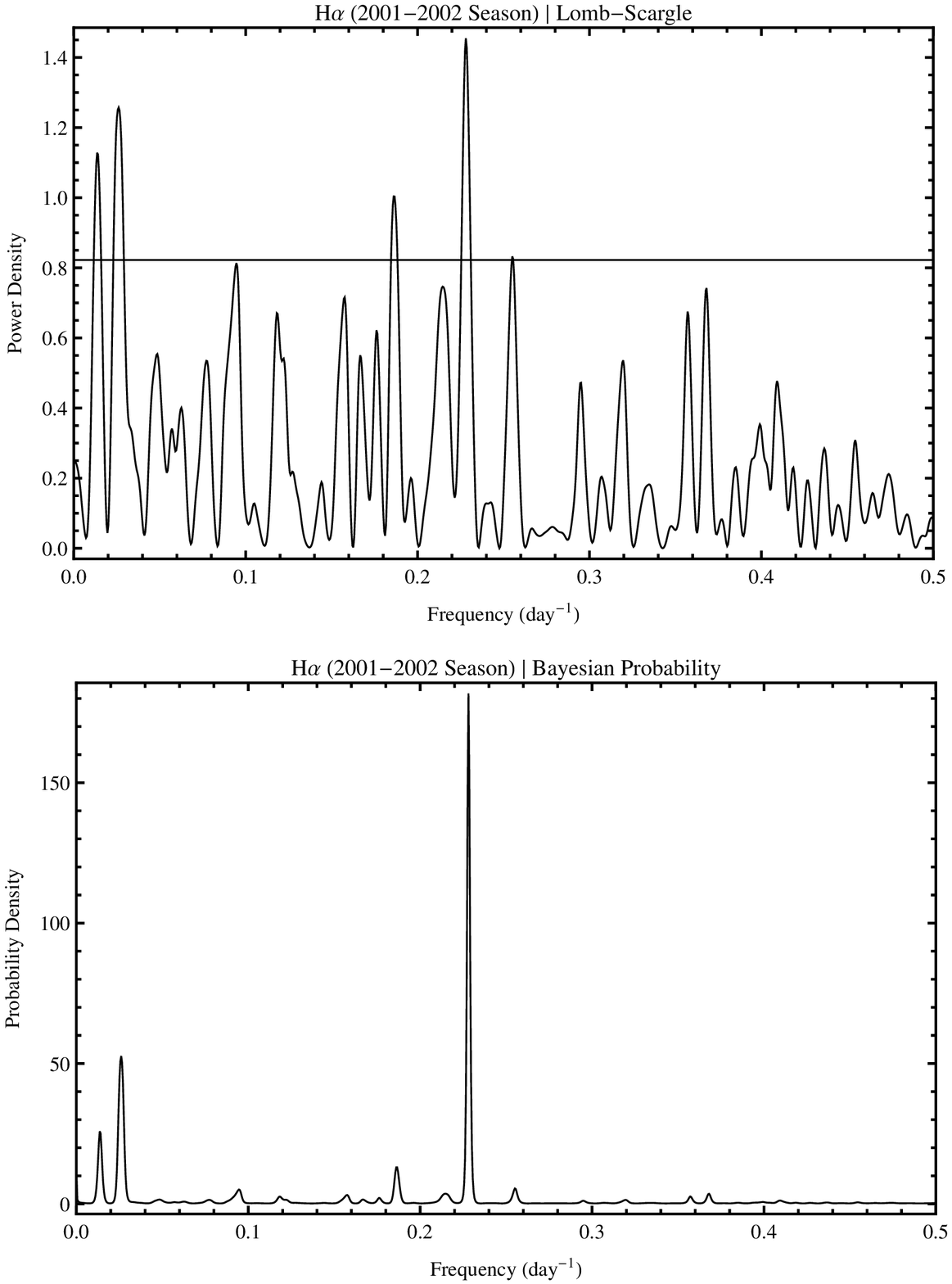}
\caption[Bayesian periodogram H$\alpha$ $W_\lambda$]{The Lomb-Scargle (upper panel) and Bayesian (lower panel) periodograms for the H$\alpha$ equivalent width time series from the 2001-2002 season.  The strongest peak locates the first frequency identified in this time series.  The horizontal line in the LS periodogram indicates the 5\% false alarm probability level.}
\label{fig7}
\end{center}
\end{figure*}

The strongest peak in the Bayesian periodogram was used to locate the first frequency identified in a time series.  A Gaussian was fit to the peak to determine the central frequency of the peak.  Once the central frequency was identified, a least-squares sine fit at that frequency was made to the data.  This fit was then used to subtract that signal from the data.  The average of the resulting secondary time series was subtracted and the analysis was repeated.  This procedure was continued through three cycles so there is a first, second, and third frequency for each time series.  With each signal subtracted the residual data set becomes more noisy.  Extending the analysis beyond three cycles was not productive.  As an example of the results, Figure \ref{fig7} shows the Lomb-Scargle and Bayesian periodograms for the H$\alpha$ equivalent width data from the 2001-2002 season.

%HaEWPer0102 fig7
  
This Bayesian algorithm was also used to calculate two-dimensional Bayesian periodograms.  Before computing the periodograms, the average spectrum for each season was subtracted.  The residual spectra were put into 1 \r{A} bins creating a time series for each wavelength bin.  The Bayesian periodogram was computed for each of these time series.  The bins covered the wavelength ranges of 6550 to 6575 \r{A} and 5868 to 5883 \r{A} for H$\alpha$ and He I, respectively.  These regions extend beyond the range of significant variability defined by the TVS for both lines.  However, only periods detected within the regions of significant variability are reported. 

\begin{figure}[htb]
\figurenum{8}
\includegraphics*[width=1.0\linewidth]{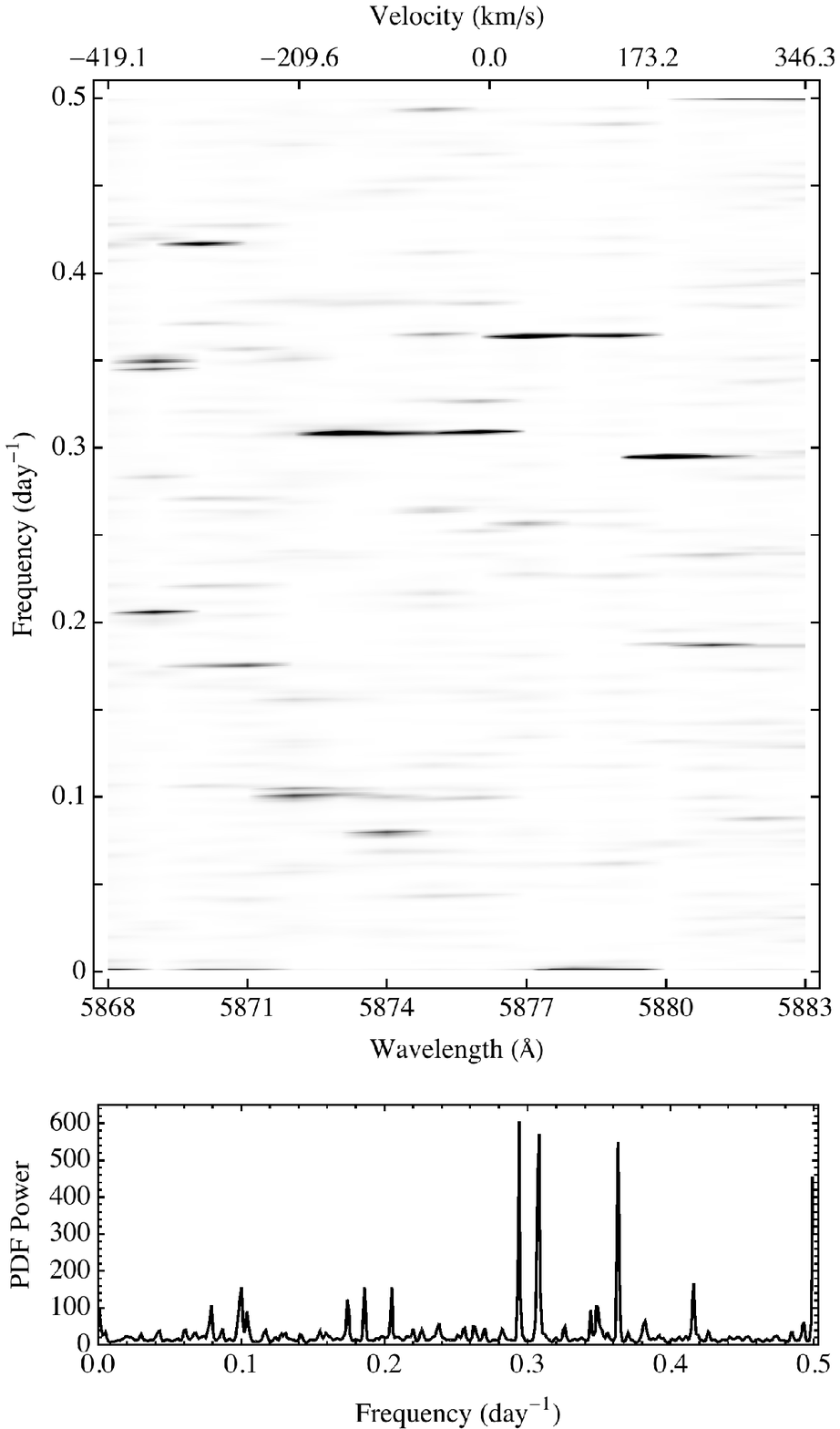}
\caption[2000-2001 He I 2D Periodogram]{The He I 2D periodogram from the 2000-2001 season.  Below the periodogram is the corresponding PDF power spectrum summed over wavelength (summed PDF power spectrum).}
\label{fig8}
\end{figure}

As an aid in identifying the strongest peaks in the 2D periodograms, the probability densities at each frequency were summed over wavelength.  For brevity, we refer to this sum as the summed PDF power as a function of frequency.  The summed PDF power was then plotted as a function of frequency to create a PDF power spectrum.  Figure \ref{fig8} shows the 2D periodogram of the He I spectra from the 2000-2001 season along with the corresponding summed PDF power spectrum.

The 2D periodograms often contained peaks at numerous frequencies.  It was desirable to limit the analysis to the strongest of these peaks.  We wanted to identify the frequencies where the most variability occurred and the wavelength bin in which that variability was the strongest.  It was found that a PDF power of 300 or more was a sufficient criterion for selecting the frequencies of strongest variability.  At this level, we were able to identify at least one wavelength bin containing a peak whose Bayesian probability was 0.2 or more for the selected frequency.  Since the power was often spread across many wavelength bins, the bin with the highest peak was chosen.  Further, this wavelength bin had to lie within the region of significant variability.

%2Dperap50001 fig8

\subsection{The CLEAN Periodogram}
A period search using CLEAN was performed on the H$\alpha$ and He I $\lambda$5876 equivalent width data.  As with the Bayesian analysis, the CLEAN algorithm was run on mean subtracted data and the search was limited to frequencies less than the Nyquist frequency.  The CLEAN analysis used the range $0 < f < 0.5$ d$^{-1}$ in frequency steps of $\delta f = 0.0005$.  The same frequency step was used to maintain consistency between the Bayesian and CLEAN approaches.  The CLEAN gain used was 0.5 and the number of iterations was set to 1000.  The central frequency of each peak was found by fitting a Gaussian to the peak.  For each frequency found, a least-squares sine fit at that frequency was made.  The fit was then used to remove that signal from the data and CLEAN was run on the new data set.  Again, this cycle was repeated three times.  

%HaEWphase0001 fig9
\begin{figure*}[]
\figurenum{9}
\begin{center}
\includegraphics*[width=0.8\linewidth]{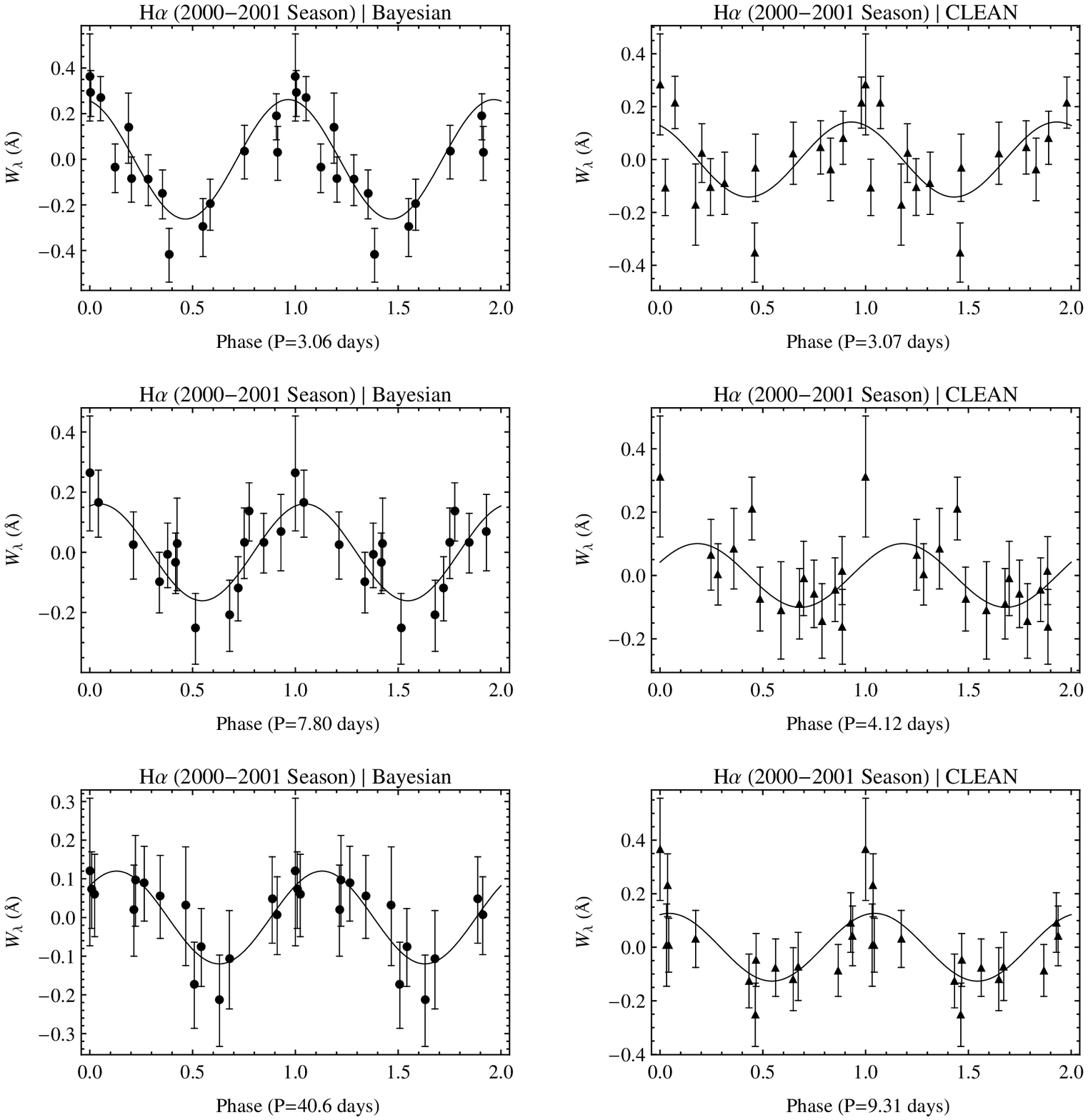}
\caption[2000-2001 H$\alpha$ Phase Plots]{Phase plots of the H$\alpha$ equivalent width data from the 2000-2001 season for the periods identified in the Bayesian and CLEAN analyses.  The mean-subtracted data were pre-whitened to remove the detected signals not being used in the current phase plot.  From top to bottom, the data are phase-folded to the first, second, and third detected periods.  For clarity, the data are plotted over two cycles with zero phase arbitrarily assigned to the first observation of each season.  The curve represents a sinusoid fit to the data.}
\label{fig9}
\end{center}
\end{figure*} 

%HeIEWphasefit9899 fig10
\begin{figure*}[]
\figurenum{10}
\begin{center}
\includegraphics*[width=0.8\linewidth]{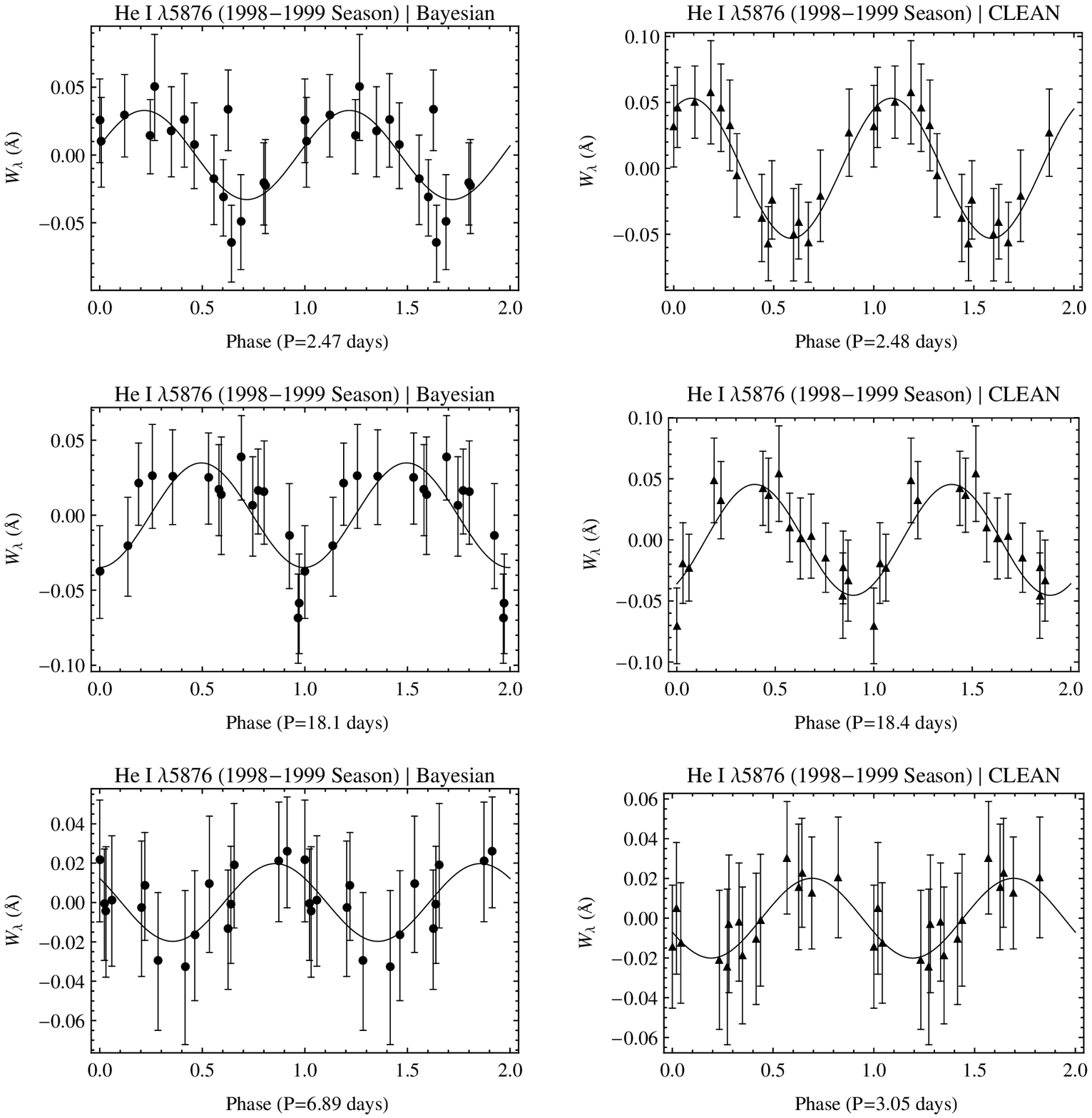}
\caption[1998-1999 He I Phase Plots]{The same as Figure \ref{fig9} but with He I $\lambda$5876 equivalent width data from the 1998-1999 season.}
\label{fig10}
\end{center}
\end{figure*}

\subsection{Evaluating the Identified Periods} \label{FAP}
Determining the significance of peaks in a periodogram is a well-known and difficult problem.  A standard metric for determining the significance of periodogram peaks is the false alarm probability (FAP).  The false alarm probability as defined by Scargle (1982) depends on a correct determination of the number of independent frequencies at which the periodogram is evaluated.  Horne \& Baliunas (1986) determined an empirical formula for calculating the number of independent frequencies, based on their Monte Carlo simulations.  Using the results of Horne \& Baliunas (1986), the 5\% false alarm probability level (or 95\% confidence) was calculated for each Lomb-Scargle periodogram.  The calculated FAP values should be viewed only as an estimate of the confidence level for each peak.  Every period reported from the Bayesian analysis had a corresponding peak in the LS periodogram that exceeded the 95\% confidence level.

A simple way of determining whether a detected period is persistent throughout a time series is to plot the data phase-folded at the period of interest.  The phase was defined as
\begin{equation}
\Phi = \frac{T-T_0}{P}-\text{Int}\left[\frac{T-T_0}{P}\right]
\end{equation}
where $T$ is the date of the observation, $T_0$ is the date representing zero phase, and $P$ is the period of interest.  If a detected signal with period $P$ is harmonic, then a sinusoidal modulation should be evident in the phase-folded data.  Phase plots were created for all detected periods.  For the equivalent width and radial velocity data, the mean-subtracted data were pre-whitened to remove the detected signals not being used in the current phase plot.  For phase plots for periods found in the 2D periodograms, the mean-subtracted flux from the bins was phase folded.

All of the frequencies detected in the analysis were converted to periods and are listed in Tables 5 and 6.  The periods from each season are listed according to data type and periodogram method.  The periods from the equivalent width and radial velocity data are listed in order of their detection (i.e. first, second, or third run through).  Periods from the 2D periodograms are simply listed in ascending order of the wavelength bin.  Tables 5 and 6 also give the Bayesian probability for the periods that were detected in Bayesian periodograms.  These probabilities were determined by simple numerical integration over the peak.  In order to provide an objective assessment of whether a sinusoidal trend is present in a phase diagram, a function of the form
\begin{equation}
A \sin(2\pi B+C)
\end{equation}
was fit to the data phased with a given period.  The parameters $A$, $B$, and $C$ were freely varied.  The correlation coefficient, $r^2$, from the fit is also given in Tables 5 and 6 for each period.  

%TABLE 5 AllPeriods Halpha
\begingroup\tabcolsep=2pt
\def\arraystretch{1.2}
\begin{deluxetable*}{cccccccccccccccccccccccc}
\tablenum{5}				
\tablecaption{\label{tab5} Detected Periods of Variability in H$\alpha$.}		
\tablewidth{500pt}	
\centering	
\tablehead{
\colhead{} &  \multicolumn{3}{c}{98-99} & \colhead{} & \multicolumn{3}{c}{00-01} & \colhead{} & \multicolumn{3}{c}{01-02} & \colhead{} & \multicolumn{3}{c}{02-03} & \colhead{} & \multicolumn{3}{c}{04-05} & \colhead{} & \multicolumn{3}{c}{05-06}\\
\cline{2-4} \cline{6-8} \cline{10-12} \cline{14-16} \cline{18-20} \cline{22-24}\\		
\colhead{Type} &
\colhead{P (d)} &
\colhead{Prob.} &
\colhead{$r^2$} &
\colhead{} &
\colhead{P (d)} &
\colhead{Prob.} &
\colhead{$r^2$} &
\colhead{} &
\colhead{P (d)} &
\colhead{Prob.} &
\colhead{$r^2$} &
\colhead{} &
\colhead{P (d)} &
\colhead{Prob.} &
\colhead{$r^2$} &
\colhead{} &
\colhead{P (d)} &
\colhead{Prob.} &
\colhead{$r^2$} &
\colhead{} &
\colhead{P (d)} &
\colhead{Prob.} &																	
\colhead{$r^2$}
}														
\startdata				
BP	&	73.4	&	0.16	&	0.60	&  &	3.06	&	0.15	&	0.76	&  &	4.38	&	0.40	&	0.40	&  &	11.2	&	0.16	&	0.64	&  &	19.3	&	0.86	&	0.63	&  &	4.32	&	0.29	&	0.56	\\
	&	4.48	&	0.12	&	0.72	&  &	7.80	&	0.13	&	0.59	&  &	41.6	&	0.26	&	0.28	&  &	4.28	&	0.50	&	0.55	&  &	4.07	&	0.13	&	0.54	&  &	2.60	&	0.12	&	0.45	\\
	&	13.1	&	0.17	&	0.52	&  &	40.6	&	0.18	&	0.72	&  &	10.7	&	0.28	&	0.23	&  &	4.66	&	0.14	&	0.38	&  &	8.65	&	0.21	&	0.35	&  &	4.39	&	0.44	&	0.45	\\
\tableline																																					
CL	&	13.4	&	\nodata	&	0.50	&  &	3.07	&	\nodata	&	0.42	&  &	38.4	&	\nodata	&	0.20	&  &	4.26	&	\nodata	&	0.59	&  &	19.1	&	\nodata	&	0.63	&  &	2.24	&	\nodata	&	0.17	\\
	&	3.51	&	\nodata	&	0.50	&  &	4.12	&	\nodata	&	0.32	&  &	4.43	&	\nodata	&	0.18	&  &	11.2	&	\nodata	&	0.71	&  &	4.70	&	\nodata	&	0.33	&  &	4.33	&	\nodata	&	0.51	\\
	&	4.30	&	\nodata	&	0.36	&  &	9.31	&	\nodata	&	0.53	&  &	11.0	&	\nodata	&	0.15	&  &	9.54	&	\nodata	&	0.36	&  &	2.77	&	\nodata	&	0.27	&  &	2.61	&	\nodata	&	0.18	\\
\tableline																																					
2D	&	13.7	&	0.74	&	0.75	&  &	2.42	&	0.25	&	0.66	&  &	4.63	&	0.40	&	0.04	&  &	21.7	&	0.74	&	0.51	&  &	31.2	&	0.46	&	0.47	&  &	5.46	&	0.41	&	0.50	\\
	&	3.62	&	0.88	&	0.69	&  &	4.55	&	0.26	&	0.66	&  &	70.9	&	0.68	&	0.48	&  &	11.2	&	0.96	&	0.59	&  &	9.70	&	0.45	&	0.47	&  &	93.0	&	0.64	&	0.58	\\
	&	2.08	&	0.89	&	0.77	&  &	3.72	&	0.22	&	0.64	&  &	5.32	&	0.55	&	0.49	&  &	2.86	&	0.64	&	0.47	&  &	6.29	&	0.49	&	0.52	&  &	6.62	&	0.81	&	0.60	\\
	&	\nodata	&	\nodata	&	\nodata	&  &	3.26	&	0.33	&	0.65	&  &	36.9	&	0.64	&	0.48	&  &	3.73	&	0.87	&	0.60	&  &	2.22	&	1.00	&	0.71	&  &	4.33	&	0.41	&	0.52	\\
	&	\nodata	&	\nodata	&	\nodata	&  &	3.06	&	0.54	&	0.71	&  &	22.7	&	0.61	&	0.45	&  &	\nodata	&	\nodata	&	\nodata	&  &	2.48	&	0.43	&	0.50	&  &	\nodata	&	\nodata	&	\nodata	\\
	&	\nodata	&	\nodata	&	\nodata	&  &	25.0	&	0.31	&	0.66	&  &	\nodata	&	\nodata	&	\nodata	&  &	\nodata	&	\nodata	&	\nodata	&  &	\nodata	&	\nodata	&	\nodata	&  &	\nodata	&	\nodata	&	\nodata				
\enddata																
\end{deluxetable*}
\endgroup

%TABLE 6 All Periods He I
\begingroup\tabcolsep=2pt
\def\arraystretch{1.2}
\begin{deluxetable*}{cccccccccccccccccccccccc}
\tablenum{6}				
\tablecaption{\label{tab6} Detected Periods of Variability in He I.}		
\tablewidth{500pt}	
\tabletypesize{\scriptsize}
\centering	
\tablehead{
\colhead{} &  \multicolumn{3}{c}{98-99} & \colhead{} & \multicolumn{3}{c}{00-01} & \colhead{} & \multicolumn{3}{c}{01-02} & \colhead{} & \multicolumn{3}{c}{02-03} & \colhead{} & \multicolumn{3}{c}{04-05} & \colhead{} & \multicolumn{3}{c}{05-06}\\
\cline{2-4} \cline{6-8} \cline{10-12} \cline{14-16} \cline{18-20} \cline{22-24}\\		
\colhead{Type} &
\colhead{P (d)} &
\colhead{Prob.} &
\colhead{$r^2$} &
\colhead{} &
\colhead{P (d)} &
\colhead{Prob.} &
\colhead{$r^2$} &
\colhead{} &
\colhead{P (d)} &
\colhead{Prob.} &
\colhead{$r^2$} &
\colhead{} &
\colhead{P (d)} &
\colhead{Prob.} &
\colhead{$r^2$} &
\colhead{} &	
\colhead{P (d)} &															
\colhead{Prob.} &																	
\colhead{$r^2$} &
\colhead{} &																
\colhead{P (d)} &																	
\colhead{Prob.} &																	
\colhead{$r^2$}
}														
\startdata
BP	&	2.47	&	0.32	&	0.56	&  &	6.39	&	0.28	&	0.88	&  &	6.01	&	0.27	&	0.58	&  &	4.64	&	0.17	&	0.00	&  &	19.4	&	0.15	&	0.67	&  &	4.23	&	0.49	&	0.61	\\
	&	18.1	&	0.94	&	0.59	&  &	5.12	&	0.12	&	0.82	&  &	4.39	&	0.19	&	0.48	&  &	4.99	&	0.85	&	0.01	&  &	4.47	&	0.21	&	0.56	&  &	4.40	&	0.41	&	0.62	\\
	&	6.89	&	0.06	&	0.53	&  &	3.15	&	0.29	&	0.71	&  &	5.28	&	0.26	&	0.40	&  &	2.10	&	0.37	&	0.02	&  &	2.67	&	0.09	&	0.37	&  &	2.65	&	0.35	&	0.46	\\
\tableline																																					
CL	&	2.48	&	\nodata	&	0.91	&  &	6.39	&	\nodata	&	0.83	&  &	6.51	&	\nodata	&	0.31	&  &	5.81	&	\nodata	&	0.18	&  &	6.26	&	\nodata	&	0.46	&  &	4.23	&	\nodata	&	0.56	\\
	&	18.4	&	\nodata	&	0.78	&  &	2.48	&	\nodata	&	0.58	&  &	3.92	&	\nodata	&	0.33	&  &	4.24	&	\nodata	&	0.11	&  &	12.0	&	\nodata	&	0.48	&  &	4.43	&	\nodata	&	0.58	\\
	&	3.05	&	\nodata	&	0.73	&  &	3.51	&	\nodata	&	0.51	&  &	5.66	&	\nodata	&	0.36	&  &	5.52	&	\nodata	&	0.06	&  &	5.71	&	\nodata	&	0.28	&  &	2.64	&	\nodata	&	0.39	\\
\tableline																																					
2D	&	3.57	&	0.86	&	0.74	&  &	3.25	&	0.38	&	0.68	&  &	39.4	&	0.43	&	0.42	&  &	2.43	&	0.76	&	0.60	&  &	4.78	&	0.35	&	0.51	&  &	3.83	&	0.56	&	0.56	\\
	&	3.10	&	0.35	&	0.74	&  &	2.75	&	0.47	&	0.69	&  &	\nodata	&	\nodata	&	\nodata	&  &	10.2	&	0.65	&	0.55	&  &	19.5	&	0.37	&	0.48	&  &	4.40	&	0.39	&	0.51	\\
	&	\nodata	&	\nodata	&	\nodata	&  &	3.40	&	0.72	&	0.75	&  &	\nodata	&	\nodata	&	\nodata	&  &	5.91	&	0.56	&	0.54	&  &	9.70	&	0.40	&	0.49	&  &	2.69	&	0.53	&	0.51	\\
	&	\nodata	&	\nodata	&	\nodata	&  &		&	\nodata	&	\nodata	&  &	\nodata	&	\nodata	&	\nodata	&  &	\nodata	&	\nodata	&	\nodata	&  &	\nodata	&	\nodata	&	\nodata	&  &	\nodata	&	\nodata	&	\nodata	\\
\tableline																																					
$V_r$	&	63.3	&	0.17	&	0.63	&  &	2.75	&	0.39	&	0.01	&  &	121	&	0.52	&	0.74	&  &	2.92	&	0.07	&	0.33	&  &	7.60	&	0.17	&	0.27	&  &	2.23	&	0.34	&	0.61	\\
	&	2.17	&	0.66	&	0.84	&  &	2.58	&	0.60	&	0.70	&  &	3.54	&	0.90	&	0.69	&  &	8.70	&	0.08	&	0.47	&  &	2.23	&	0.50	&	0.70	&  &	5.00	&	0.52	&	0.60	\\
	&	16.2	&	0.69	&	0.69	&  &	2.89	&	0.28	&	0.64	&  &	3.85	&	0.21	&	0.38	&  &	9.86	&	0.10	&	0.36	&  &	3.33	&	0.69	&	0.50	&  &	5.90	&	0.73	&	0.58	
\enddata																	
\end{deluxetable*}
\endgroup

\begin{figure}[]
\figurenum{11}
\begin{center}
\includegraphics*[width=0.9\linewidth]{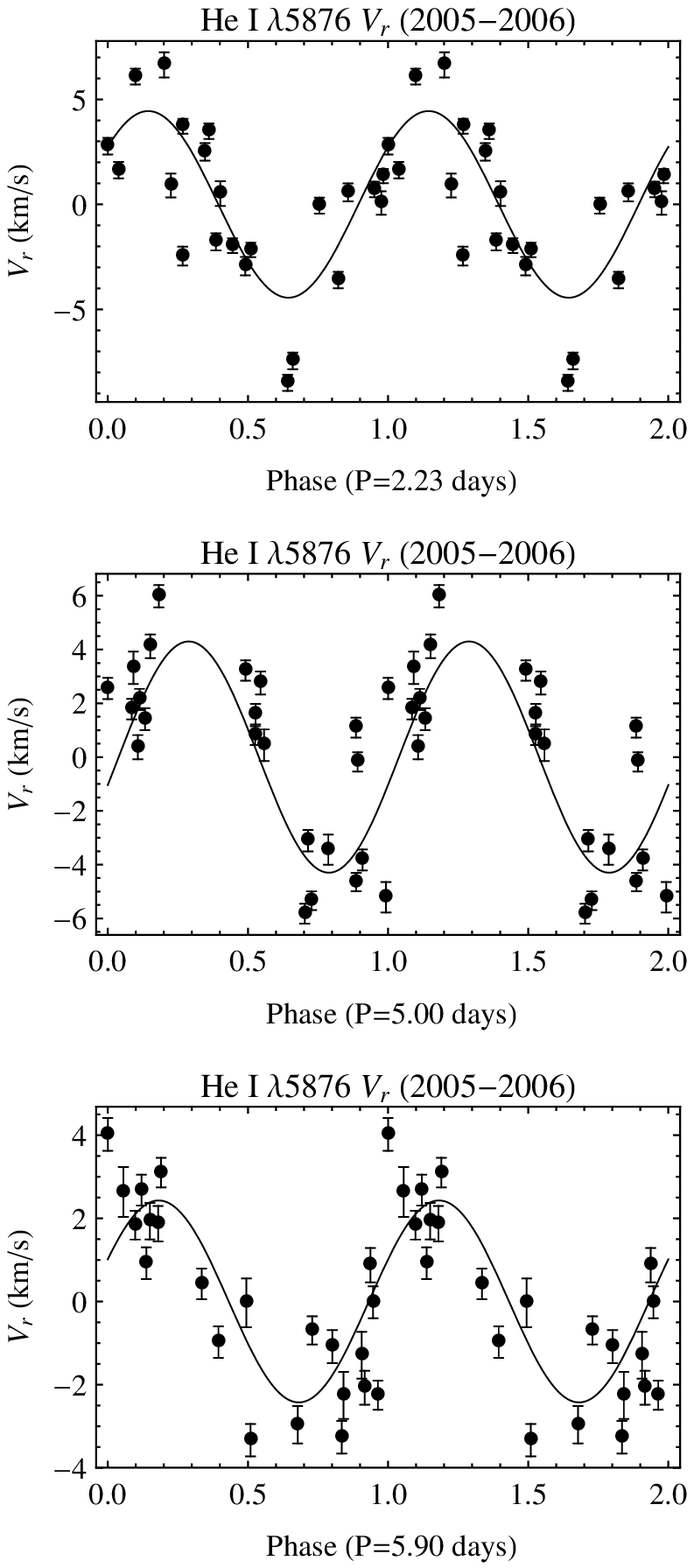}
\caption[2005-2006 He I $V_r$ Phase Plots]{The same as Figure \ref{fig9} but with He I $\lambda$5876 radial velocity data from the 2005-2006 season.}
\label{fig11}
\end{center}
\end{figure}

As examples of the phase diagrams, Figures \ref{fig9} - \ref{fig11} show the phase diagrams for periods found in the 2000-2001 H$\alpha$ and 1998-1999 He I equivalent width data and the radial velocities from the 2005-2006 data.  For clarity, the phase diagrams are plotted over two cycles with zero phase arbitrarily set to the first observation of the season.  In Fig. \ref{fig9}, the phase diagram of the \mbox{3.06 d} period found in the H$\alpha$ equivalent width data is reasonably well fit by the sinusoid with an $r^2$ value of 0.76.  In Fig. \ref{fig10}, the \mbox{2.48 d} period found in the He I equivalent width data is fit very well, with an $r^2$ value of 0.91 while the 18.4 d period had an $r^2$ value of $\sim$0.78.  We found that an $r^2$ value of $\sim$0.7 or more was generally a good indication that sinusoidal modulation was present.  However, there were a few cases where modulation was judged by eye to be present even with an $r^2$ value as low as 0.48.  From these figures, it is also clear that the Bayesian and CLEAN methods often find the same periods, though not always in the same order.  

%HeIRV0506phasefit fig11

\begin{figure*}[]
\figurenum{12}
\begin{center}
\includegraphics*[width=0.8\linewidth]{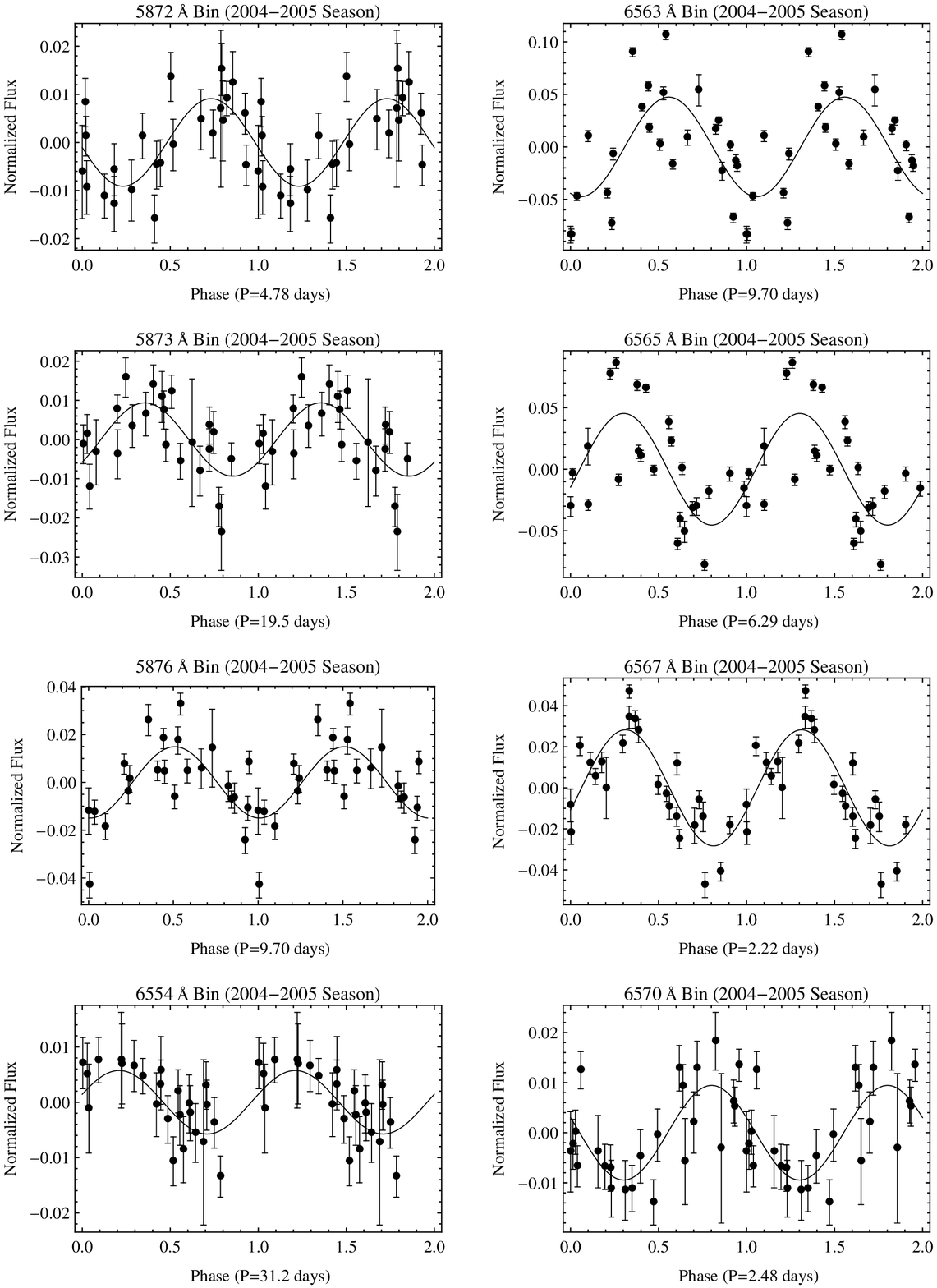}
\caption[Phase Plots From 2004-2005 2D Periodograms]{Phase plots of the wavelength bins containing the frequencies detected in the 2D periodograms from the 2004-2005 season.  Notation is as in Figures 9 - 11 .}
\label{fig12}
\end{center}
\end{figure*}

Phase diagrams were also made for periods detected in the 2D periodograms.  Here, the wavelength bin that contained the peak was phase-folded at the particular period.  In many cases a frequency was detected over several wavelength bins but only the bin with the highest peak was used for the phase plot.  The error bars on the data points represent $\sigma/2$, where $\sigma$ is the reciprocal of the S/N in the continuum of each spectrum.  The phase plots from the 2004-2005 season are shown in Fig. \ref{fig12}.  In this case, the 19.5 and 9.7 d periods appear to show sinusoidal modulation even though their $r^2$ values are $\sim$0.5. 

%ph04052Da fig12

%##########################################################################

\section{RESULTS OF THE PERIOD SEARCH} \label{sec7}
In assessing the many periods found in the period search, we used several criteria.  First, periods identified using more than one technique were considered more robust detections.  Second, periods detected in both H$\alpha$ and He I were of interest because of the possible wind-photosphere connection.  Third, we looked for sinusoidal modulation in the phase diagrams.  Finally, periods that were on the order of the rotational period of the star (13-18 d; see Sec. 7.2 below) were of interest because they could indicate rotational modulation.  Some key results are listed in Table 7.

%TABLE 7
\begingroup\tabcolsep=3pt
\begin{deluxetable*}{lcccccclc}[]
\tablenum{7}
\tablecaption{\label{tab7} Key Results of the Period Search.}
\tablewidth{400pt}
\tabletypesize{\footnotesize}
\tablehead{
  \colhead{Season} &
  \colhead{} &
  \colhead{Period} &
  \colhead{H$\alpha$ $W_\lambda$} &
  \colhead{He I $W_\lambda$} &
  \colhead{ $V_r$ } &
  \colhead{2D Per.} &
  \colhead{Bin(s)} &
  \colhead{Modulation in phase diagram?} \\
  \colhead{} &
  \colhead{} &
  \colhead{(d)} &
  \colhead{} &
  \colhead{} &
  \colhead{} &
  \colhead{} &
  \colhead{(\r{A})} &
  \colhead{} }    
\startdata
1998$-$1999	&&	2.47, 2.48	&		&	$\checkmark$	&		&		&		&	yes	\\
	&&	3.51$-$3.62	&	$\checkmark$	&		&		&	$\checkmark$	&	5876, 6566	&	yes	\\
	&&	13.1$-$13.7	&	$\checkmark$	&		&		&	$\checkmark$	&	6562	&	yes	\\
	&&	18.1, 18.4	&		&	$\checkmark$	&		&		&		&	yes	\\
\tableline															
2000$-$2001	&&	2.75	&		&		&	$\checkmark$	&	$\checkmark$	&	5877	&	yes	\\
	&&	3.06, 3.07	&	$\checkmark$	&		&		&	$\checkmark$	&	6564	&	yes	\\
	&&	3.25, 3.26	&		&		&		&	$\checkmark$	&	5876, 6561	&	yes	\\
	&&	6.39	&		&	$\checkmark$	&		&		&		&	yes	\\
\tableline															
2001$-$2002	&&	4.38$-$4.43	&	$\checkmark$	&	$\checkmark$	&		&		&		&	no	\\
\tableline															
2002$-$2003	&&	4.24$-$4.28	&	$\checkmark$	&	$\checkmark$	&		&		&		&	yes	\\
	&&	11.2	&	$\checkmark$	&		&		&	$\checkmark$	&	6563	&	yes	\\
\tableline															
2004$-$2005	&&	2.22, 2.23	&		&		&	$\checkmark$	&	$\checkmark$	&	6567	&	yes	\\
	&&	9.7	&		&		&		&	$\checkmark$	&	5876, 6563	&	yes	\\
	&&	19.1$-$19.5	&	$\checkmark$	&	$\checkmark$	&		&	$\checkmark$	&	5873	&	yes	\\
\tableline															
2005$-$2006	&&	2.23, 2.24	&	$\checkmark$	&		&	$\checkmark$	&		&		&	yes	\\
	&&	2.60$-$2.69	&	$\checkmark$	&	$\checkmark$	&		&	$\checkmark$	&	5878	&	no	\\
	&&	4.32$-$4.43	&	$\checkmark$	&	$\checkmark$	&		&	$\checkmark$	&	5876, 6568	&	yes	
\enddata
\tablecomments{In cases where three or more similar periods were identified within a season the range is given.}
\end{deluxetable*}
\endgroup

Many of the periods detected in our search were in the \mbox{2 - 7 d} range with no obvious connection to rotational modulation.  These periods may be associated with stellar pulsations or some type of evanescent events.  In the He I $W_\lambda$ data from 2000-2001, a period of \mbox{6.39 d} was identified in by both Bayesian and CLEAN periodograms, with sinusoidal modulation quite strong in both phase diagrams.  A \mbox{6.89 d} period was found in the He I $W_\lambda$ data of the 1998-1999 season and periods of 6.01 and \mbox{6.51 d} were detected in the He I $W_\lambda$ data of the 2001-2002 season.  In addition, a \mbox{6.62 d} period was found in the 6565 \r{A} bin of the 2005-2006 spectra.  These periods are all similar to the \mbox{6.6 d} period reported in spectra from 1998 by Prinja et al. (2004) and may be due to a persistent or recurring phenomenon.

\subsection{Evidence For A Wind-Photosphere Connection}
Many of the identified periods, especially those in the H$\alpha$ equivalent width, showed harmonic modulation in their phase diagrams.  This modulation is particularly associated with periods found in the H$\alpha$ $W_\lambda$.  Given the scaling relation found by Puls et al. (1996) between H$\alpha$ $W_\lambda$ and the mass-loss rate, it is tempting to conclude that the star's mass-loss rate is undergoing harmonic modulation, possibly due to stellar pulsations.  However, caution is warranted in making any such connection.  If the wind is translucent, the variable flux from pulsations at the surface may be modulating the ionization fraction of hydrogen, leading to changes in the H$\alpha$ equivalent width.

In the 2004-2005 season, a period of \mbox{$\simeq$2.2 d} was found in the radial velocity data as well as in the 6567 \r{A} bin of the 2D periodogram.  This period was also detected in the H$\alpha$ $W_\lambda$ and radial velocity data in the 2005-2006 season.  Harmonic modulation was present in the phase diagrams for these periods.  The fact that this period was found in back-to-back seasons implies that this may be a persistent or recurrent signal.  Indeed, in the 1998-1999 season a period of \mbox{2.17 d} was detected in the radial velocity data and a \mbox{2.08 d} period was found in the 6572 \r{A} bin of the 2D periodogram.  

Inspection of Tables 4 - 7 reveals that periods of \mbox{4.3 - 4.6 d} were detected in every season.  Sinusoidal modulation was evident in some of the phase diagrams for these periods.  These periods were found in the equivalent width data as well as the 2D periodograms but not in the radial velocity data.  It is interesting to note that the \mbox{2.2 d} period may be an overtone of the \mbox{$\simeq$4.4 d} periods.  If these periods are separate detections of the same signal, it indicates a persistent or recurrent trend.    

From the 2D periodograms, we find that much of the variability in He I is detected near the central wavelength bin suggesting a photospheric origin.  Meanwhile, this same period may be detected in H$\alpha$ but in a bin that implies an origin in the wind.  For example, in 1998-1999 spectra a period of \mbox{3.57 d} was found in the 5876 \r{A} bin and a period of \mbox{3.62 d} was found in the 6566 \r{A} bin.  In the 2005-2006 data, a period of \mbox{4.40 d} was identified 5876 \r{A} bin and a period of \mbox{4.40 d} was found in the 6568 \r{A} bin.

\subsection{Possible Rotational Modulation}
The time scales of the periods can provide a hint as to the origin of the variability.  A period due to rotational modulation of wind structure, such as a co-rotating interaction region (CIR), would be expected to be on the same time scale as the rotational period of the star.  Rotational modulation could also have time scales that are some fraction of the rotation period, if there were more than one wind structure present and they were approximately equally spaced.  The dependence of the rotational period on the radius of the star is a major source of uncertainty.  An accurate angular diameter measurement could help to constrain the stellar radius and the rotational period.  The projected rotational velocity is another source of uncertainty since it is an upper limit, and a fraction of it may be due to macroturbuence rather than rotation.  Adopting a $v\sin{i}$ of 91 km s$^{-1}$ and a stellar radius of 32 R$_\odot$ gives a maximum rotational period of $\simeq$18 d.  Assuming a radius of 24 R$_\odot$ with the same $v\sin{i}$ yields a maximum rotational period of $\simeq$13 d.  

Several of the detected periods fall within the range of \mbox{13-18 d} estimated for the period of rotation.  In the He I $W_\lambda$ data from the 1998-1999 season periods of 18.1 and \mbox{18.4 d} were found by the Bayesian and CLEAN periodograms, respectively.  Both phase diagrams exhibited sinusoidal modulation, particularly the \mbox{18.4 d} period.  A period of \mbox{16.2 d} was detected in the radial velocity data from the same season.  It is possible that these two periods have the same origin but with the \mbox{16.2 d} period rooted in the photosphere while the \mbox{18 d} period arises farther out in the wind.  

Periods ranging from 19.1 to \mbox{19.5 d} were identified in both the H$\alpha$ and He I $W_\lambda$ data for the 2004-2005 season.  The phase diagrams for these periods all exhibit sinusoidal modulation.  A period of \mbox{19.5 d} was also found in the 5873 \r{A} bin of the 2D periodogram the same season.  The 5873 \r{A} bin is in the wavelength region where we begin to see wind contamination in the He I line.  Also in this season, a period of \mbox{9.7 d} was found in the central wavelength bins of 5876 and 6563 \r{A} in the 2D periodograms.  The \mbox{9.7 d} period could be an overtone of the \mbox{$\simeq$19.4 d} periods and may indicate a connection between the photosphere and wind emission.  A \mbox{9.7 d} period was also reported by Prinja et al. (2004) in H$\alpha$, H$\beta$, and He I $\lambda$6678 observations from 1998.

%##########################################################################

\section{SUMMARY AND CONCLUSIONS} \label{sec8}

The line profile variability of the H$\alpha$ and He I $\lambda$5876 lines in the supergiant $\epsilon$ Ori has been examined through time-series analysis of spectra from seven seasons of observations.  This star has a strong and variable stellar wind, as demonstrated by its highly variable H$\alpha$ line which is observed to have several profile morphologies including P Cygni, inverse-P Cygni, double absorption, and pure emission.  Profiles of the H$\alpha$ line show net emission approximately 65\% of the time.  The He I line is primarily an absorption line with weak wind features in the wings of the line.  The He I profile is seen to sway back and forth, possibly indicating pulsational radial velocity oscillations.    

Temporal variance spectra were calculated for each season for both spectral lines.  The TVS clearly show that both lines undergo significant variability extending beyond $\pm$$v \sin{i}$ region indicating the presence of wind variability.  In both cases, the variability is greatest on the blue side of line center.  The TVS for the He I line has a double-peaked profile indicative of radial velocity oscillations due to pulsations.  The double peaks are contained within the $v \sin{i}$ region consistent with a photospheric or near photospheric origin for the radial velocity variations.  The region of significant variability defined by the 99\% confidence level of the TVS was used to set the wavelength regions over which equivalent width measurements were made for the H$\alpha$ and He I spectra.  Radial velocity measurements were also made from the central absorption of the He I profile.  

A periodicity search was carried out on data from six of the seven seasons individually.  A technique that applies Bayesian statistics to the Lomb-Scargle periodogram was employed on the equivalent width, radial velocity, and binned-spectra time series.  The CLEAN algorithm was also used on the equivalent width time series.  Numerous periods were identified in the variability as a result of these analyses.  All periods identified in the Bayesian analysis had corresponding LS periodogram peaks that exceeded the 5\% false alarm probability.  Phase diagrams were created for each identified period and a sinusoidal function was fit to the phased data.  The resulting phase plots indicate that several of the periods are harmonic.

Periods on the order of the rotational period of the star (13-18 d) were found in data from two seasons indicating possible rotational modulation.  Sinusoidal modulation was evident in the phase diagrams in both cases.  In the 1998-1999 observing season, periods \mbox{$\simeq$18 d} were detected in the He I equivalent width data.  In the 2004-2005 H$\alpha$ and He I equivalent width data, periods of 19.1 to \mbox{19.5 d}  were found.  A \mbox{9.7 d} period was also present in the 5876 and 6563 \r{A} bins in the 2D periodograms from that season.  This is a possible overtone of the \mbox{$\sim$19 d} period.

Most of the periods identified were in the range of \mbox{2 - 7 d}.  Many of these periods were present in both H$\alpha$ and He I data and exhibited sinusoidal patterns in their phase diagrams.  Stellar pulsations are a possible origin of this variability and provide evidence for a connection between the photospheric activity and the wind.  Particularly intriguing is the fact that periods of \mbox{4.3 - 4.6 d} are present in every season and were found in both lines.  This may represent a persistent mode of variability in this star. 

The spectral variability of $\epsilon$ Ori is clearly complex and in need of further study.  Stellar pulsation modeling of this star is needed in order to determine if the periods reported here can be produced by radial or non-radial pulsations.  Observationally, a campaign to acquire simultaneous spectroscopic and photometric data could provide conclusive evidence for a connection between stellar pulsation and wind variability.  A more complete time series at optical and UV wavelengths, probing the inner and outer wind regions, would provide a more complete picture of the wind variability.

%##########################################################################

\begin{acknowledgments}

We wish to thank Jon Bjorkman, Karen Bjorkman, Douglas Gies, and Thomas Kvale, for their support and many excellent suggestions.  This work was partially supported by an NSF-PREST Grant (AST-0440784) awarded to Nancy Morrison and Karen Bjorkman.  Support for the acquisition of these spectra at Ritter Observatory came from: The University of Toledo, The Fund for Astrophysical Research, Inc., and the American Astronomical Society Small Research Grants Program.  Financial support for the publication of this work was received from the Scott E. Smith Fund for Research at Ritter Observatory and from Adrian College.  Finally, thanks go to the following members of the Ritter observing team for obtaining spectra of $\epsilon$ Ori:  Nancy Morrison, Chris Mulliss, David Knauth, Howard Ritter, Will Fischer, Karen Bjorkman, Anatoly Miroshnichenko, John Wisniewski, Josh Thomas, Amanda Gault, Doug Long, Noel Richardson, Erica Hesselbach, Nick Sperling, Dan Kittell, and Ian McGinness.

\end{acknowledgments}

{\it Facility:} \facility{Ritter Observatory}

%========================================

%##########################################################################
\end{document}